# A single coordinate framework for optic flow and binocular disparity

Andrew Glennerster[1¶*] and Jenny C.A. Read[2¶]

[1] School of Psychology and Clinical Language Sciences,
University of Reading,
Reading RG6 6AL

[2] Institute of Neuroscience,
Newcastle University,
NE2 4HH

[¶] These authors contributed equally to this work.

[*] Corresponding author. Email a.glennerster@reading.ac.uk (AG)

## Abstract

Optic flow is two dimensional, but no special qualities are attached to one or other of these dimensions. For binocular disparity, on the other hand, the terms 'horizontal' and 'vertical' disparities are commonly used. This is odd, since binocular disparity and optic flow describe essentially the same thing. The difference is that, generally, people tend to fixate relatively close to the direction of heading as they move, meaning that fixation is close to the optic flow epipole, whereas, for binocular vision, fixation is close to the head-centric midline, i.e. approximately 90 degrees from the binocular epipole. For fixating animals, some separations of flow may lead to simple algorithms for the judgement of surface structure and the control of action. We consider the following canonical flow patterns that sum to produce overall flow: (i) '*towards' flow*, the component of translational flow produced by approaching (or retreating from) the fixated object, which produces pure radial flow on the retina; (ii) *'sideways' flow*, the remaining component of translational flow, which is produced by translation of the optic centre orthogonal to the cyclopean line of sight and (iii) *'vergence' flow,* rotational flow produced by a counter-rotation of the eye in order to maintain fixation. A general flow pattern could also include (iv) '*cyclovergence' flow*, produced by rotation of one eye relative to the other about the line of sight. We consider some practical advantages of dividing up flow in this way when an observer fixates as they move. As in some previous treatments, we suggest that there are certain tasks for which it is sensible to consider 'towards' flow as one component and 'sideways' + 'vergence' flow as another.

## Author Summary

"Optic flow" refers to changes in the visual images we receive as we move through a scene. For example, as we drive along a street, the buildings flow past us and distant objects expand in our field of view. This information can tell us both about how we are moving and about the 3D structure of the scene. Conversely, "binocular disparity" refers to the differences between the views seen by our two eyes due to their slightly different positions in their head. Binocular disparity enables us to perceive the distance to objects, even while stationary. Both flow and disparity are rich sources of information, both for humans and other animals, and for robot systems such as autonomous drones and cars. Generally, they have been studied separately, and a separate set of vocabulary has been developed for each. Yet mathematically the two are fundamentally related: optic flow compares views seen by a single, moving eye at two different points in time, while binocular disparity compares views seen simultaneously by two eyes at different positions. Here, we develop a common language for describing both. It is particularly appropriate for eyes that fixate as they move, a universal feature of biological visual systems.

## 1. Introduction

According to the ancient Indian parable, a number of blind men examined an elephant and, because each came into contact with a different part, such as the tusk or the tail, they came up with entirely different conclusions about the nature of the beast (Saxe, 1881). The same could be said about vision researchers' approaches to optic flow and binocular disparity, where the 'elephant' is the flow field generated on the retina by movement of the eye through space (or between the left and right eyes).

As an illustration, imagine a train travelling along a track (Figure 1). The (monocular) train driver, looking through the windscreen, sees the scene expanding from a single point in the image (Figure 1, right) while a (monocular) passenger, looking out through a window, sees the landscape moving sideways (Figure 1, left). Yet these two very different flow fields are in fact both samples from the same spherical flow field (Figure 1, centre). The train driver is looking straight ahead at the 'direction of heading' or 'focus of expansion'. This is known as the *epipole,* i.e. the part of the image pierced by the translation vector. The passenger samples a quite different part of the flow field and, for him or her, the epipole is obscured from view.

In neuroscience, those studying optic flow concentrate on the driver's view and those studying binocular stereopsis on the passenger's.  In the case of binocular vision, the translation of the optic centre that generates flow is the interocular separation and the epipole is always obscured from view. In this case, no rays can reach the eye along the interocular axis, just like the passenger in the train who cannot see the place that his or her carriage is heading. Different nomenclatures have arisen in the two communities; for example, 'flow' is known as 'disparity' in binocular vision. However, there is no good reason to use different terms and it would be helpful to have a general framework that encompasses both. That is the goal of this paper.

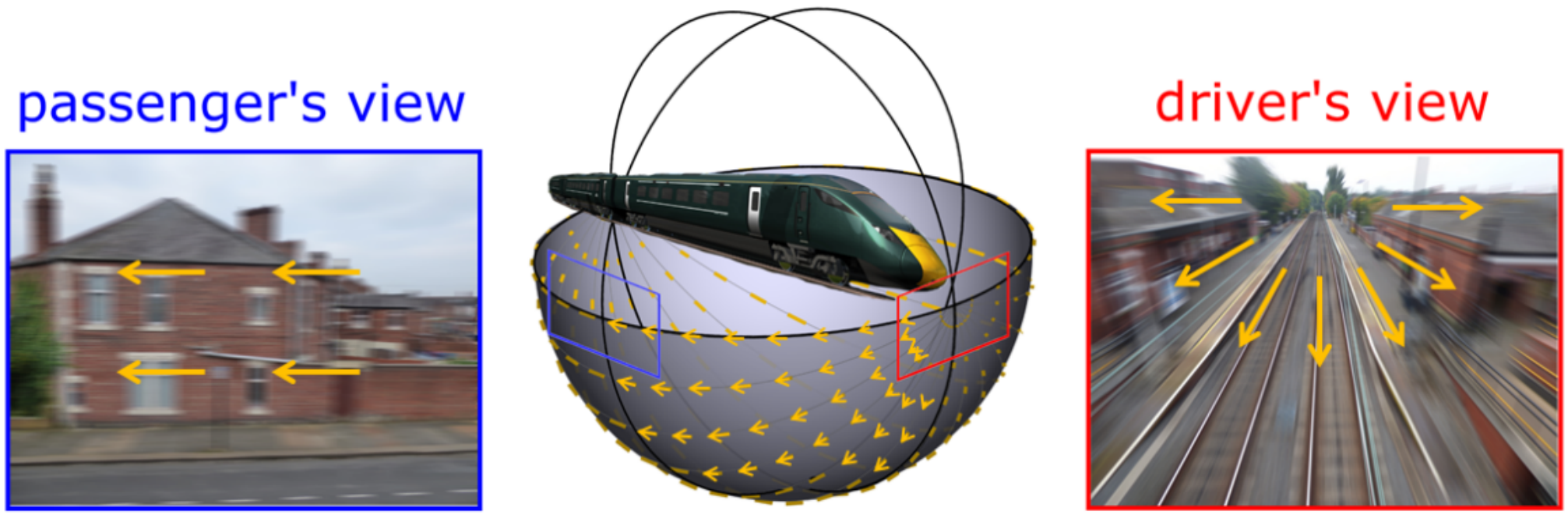

***Figure 1. The pattern of flow produced as a train travels in a straight line.*** *The orange arrows mark the flow lines on the visual sphere. For a passenger looking out of a side window, the flow is approximately horizontal ('sideways' flow); for the driver, the flowfield is radial ('towards' flow).*

One apparent point of difference between binocular disparity and optic flow is the length of the baseline. The eyes are separated by a 'baseline' distance of about 6.5cm, whereas for optic flow the baseline separating the two locations of the optic centre at time 1 and time 2 is typically considered to be much smaller, even infinitesimal. However, flow in general (i.e. optic flow or binocular disparity) is determined by the distance of the object relative to the length of the baseline, so any statements that are applicable to a wide range of object distances are applicable for both a large and a small baseline. Also, the 3D structure of the world could change between time 1 and time 2, whereas for binocular stereopsis the views are always simultaneous. However, in this paper we consider only a moving observer (or eye) in a static 3D world so that binocular vision and optic flow are equivalent in this regard. Note that we are *not* considering the integration of disparity and motion information as a binocular observer moves (Cormack et al, 2017). We are considering a unified framework for describing flow across the whole sphere, both near the epipole (traditionally, the part of the sphere that is considered by optic flow researchers) and around 90 degrees from the epipole (traditionally, the part of the sphere that is considered by binocular vision researchers).

Examples of terms that are used in the binocular vision literature with no direct analogue in the optic flow literature are: horizontal disparity, vertical disparity, polar angle disparity and horizontal or vertical size ratios. Optic flow uses some terms that are not used in binocular vision such as 'div', 'def', and 'curl' components (Koenderink and van Doorn, 1976) or the 'centre of the expanding flow pattern' (Regan and Beverley, 1982) although these are less commonly used now. By contrast, the computer vision photogrammetry literature simply defines two basis vectors, usually based on the 2D pixel grid, to describe the change in image location of features (Hartley and Zisserman, 2003). There is no special or different treatment of flow in a 'horizontal' or 'vertical' direction – all flow is treated equally. The question we explore in this paper is whether there is any compelling theoretical rationale for dividing retinal flow up in particular ways and attaching special meanings to different components (such as 'horizontal' or 'vertical'). An important factor will be the constraints that are introduced by the eye movements humans make, especially the fact that observers maintain fixation on a point as they move or when they view a scene binocularly. We begin by giving an intuitive overview of the main arguments in the paper, followed by a more formal description (Section 2.2).

# 2. Results

## 2.1. Overview: decomposing flow in a fixating system

We begin by describing in outline the key contributions of the paper. Figure 2A illustrates two eyes (or a single eye at time 1 and time 2) fixating a point *F*. In biology, most animals fixate as they move (Land, 1999; Land and Nilsson, 2012). This means that as the eye translates from $O_1$ to $O_2$, it also rotates so as to keep the visual axis fixated on *F*. For simplicity, we also assume that cycloversion of the eye is minimized with respect to the scene, which would mean that the horizon projects to the same retinal meridian before and after the translation (for a review of torsional eye movements during head translation, see Angelaki et al, 2003). This yoking of rotation to translation reduces the six degrees of freedom of the camera/eye (3 translation and 3 rotation) to just three. So, in the simple case illustrated here, optic flow is the sum of just two components: translational flow caused by the displacement of the optic centre (Figure 2B) and rotational flow caused by the rotation of the visual axis necessary to maintain fixation (Figure 2C). We refer to this rotational flow as 'vergence flow', using the terminology from binocular vision. As shown in Figure 2, translational flowlines are lines of longitude on the retina, with a pole corresponding to the direction of heading (the 'epipole'). In binocular vision, the 'direction of heading' is along the interocular axis. Rotational flowlines are lines of latitude on the retina, with a pole corresponding to the axis of rotation. For a fixating eye, this is orthogonal to the plane of regard (the plane $O_1O_2F$ shown in Figure 2A), so the plane of regard corresponds to the 'equator' of the rotational flowlines.

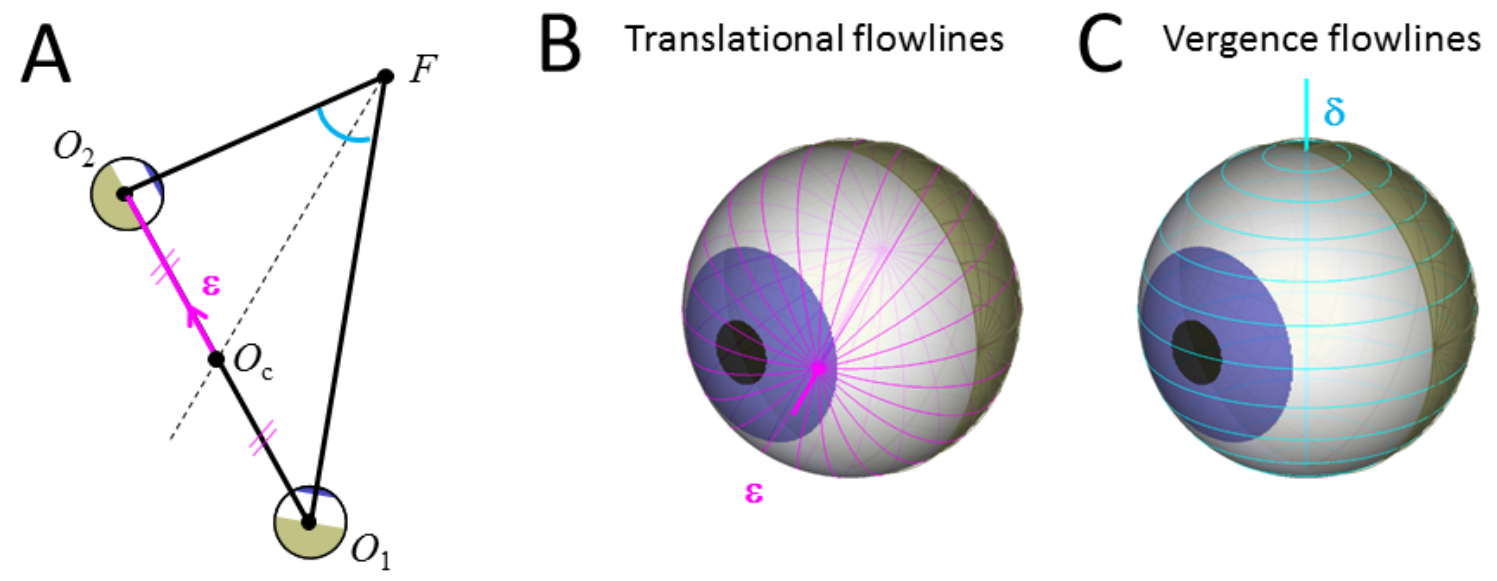

***Figure 2. Translational and rotational components of flow.*** *For an observer maintaining fixation on a point F, translation from optic centre $O_1$ to $O_2$ has to be accompanied by a vergence rotation about an axis normal to plane $O_1O_2F$. Here, we assume cyclovergence is zero. (***A***): top-down view of the eyes and the fixation point. $O_c$ is the optic centre of the cyclopean eye, defined to be midway in between Eye 1 and Eye 2. The plane containing $O_1$, $O_2$ and F is called the 'plane of regard'. (***B***): Flowlines caused by a pure translation along the epipolar vector,* ***ε****. These are lines of longitude with poles defined by* ***ε****. (***C***): Flowlines caused by a pure vergence rotation, about an axis* ***δ*** *normal to the plane containing the optic centres and the fixation point. These are lines of latitude with poles defined by* ***δ****.*

In this paper, we will argue that it is advantageous to further divide the translation into two: a vector **t** generated by translation of the optic centre towards the fixation point, *F,* and a vector **s** generated by translation of the optic centre in an orthogonal direction, i.e. sideways (Figure 3A). These components generate flow patterns we refer to as 'towards' and 'sideways' flow. For example, approaching the fixated object results in 'towards' flow (Figure 3C) while side-to-side 'bobbing' head movements to obtain distance estimates from motion parallax would result in pure 'sideways' flow (Figure 3D) if the observer were fixating a distant point so that there was no rotational flow. General translational flow (Figure 3B) can be expressed as the sum of these two flow patterns.

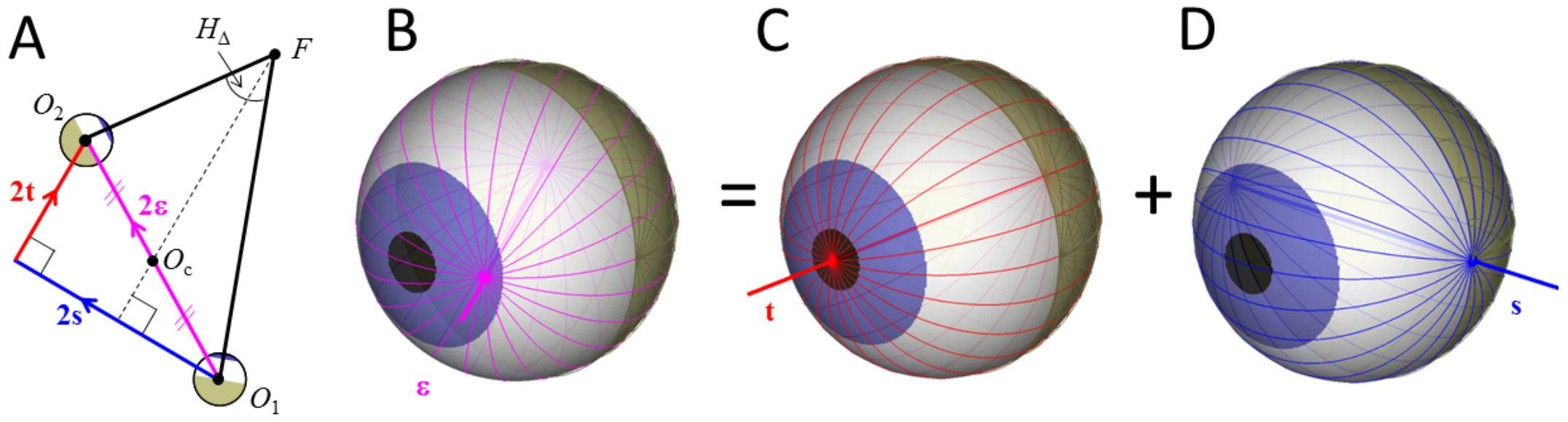

***Figure 3. Decomposition of translational flow.*** *A) The epipolar vector,* ***ε****, can be decomposed into two components,* ***s*** *('sideways') and* ***t*** *('towards'), where* ***t*** *is parallel to the line $O_cF$ and $O_c$ is midway between $O_1$ and $O_2$. The factor of 2 is included for consistency with later notation. The same is true for flow: general translational flow (**B**) can be decomposed into 'towards' flow (**C**) and 'sideways' flow (**D**).*

This type of decomposition is familiar in the optic flow literature for small patches of the image. For example, in a static scene the 'divergence' component of flow for a local surface patch is generated by translation of the optic centre towards the surface (Koenderink and van Doorn, 1976; Koenderink, 1986), i.e. 'towards' flow. However, exactly the same geometry also applies to binocular vision (Figure 4). When the fixated object is on the head-centric midline (Figure 4A) the 'towards' component of translation is zero and the 'sideways' component is equal to the interocular separation. The disparity vectors are all along horizontal lines of longitude, as in the 'sideways' flowfield of Figure 3D. When an observer views an object that is 20 degrees to the left of the headcentric midline (Figure 4B), the sideways component of translation is now slightly smaller than the interocular separation and there is an additional 'towards' component of translation, leading to an expansion of the image in the left eye relative to the right. Consequently, disparities now also have components of the radial ‘towards’ flowfield shown in Figure 3C. If the observer now rotates his or her head about a vertical axis while maintaining fixation, so that the object is 20 degrees to the *right* of their midline, the ‘sideways’ component of translation is unchanged but the ‘towards’ component changes sign, so that this component now results in a relative expansion of the fixated object in the *right* eye (Figure 4C) compared to the left eye. In general, therefore, the ‘sideways’ component of disparity provides a useful signal about the shape of the surface while the ‘towards’ component provides a signal relating to the position of the surface relative to the midline (Backus et al (1999), Rogers and Bradshaw (1993) and Section 2.4).

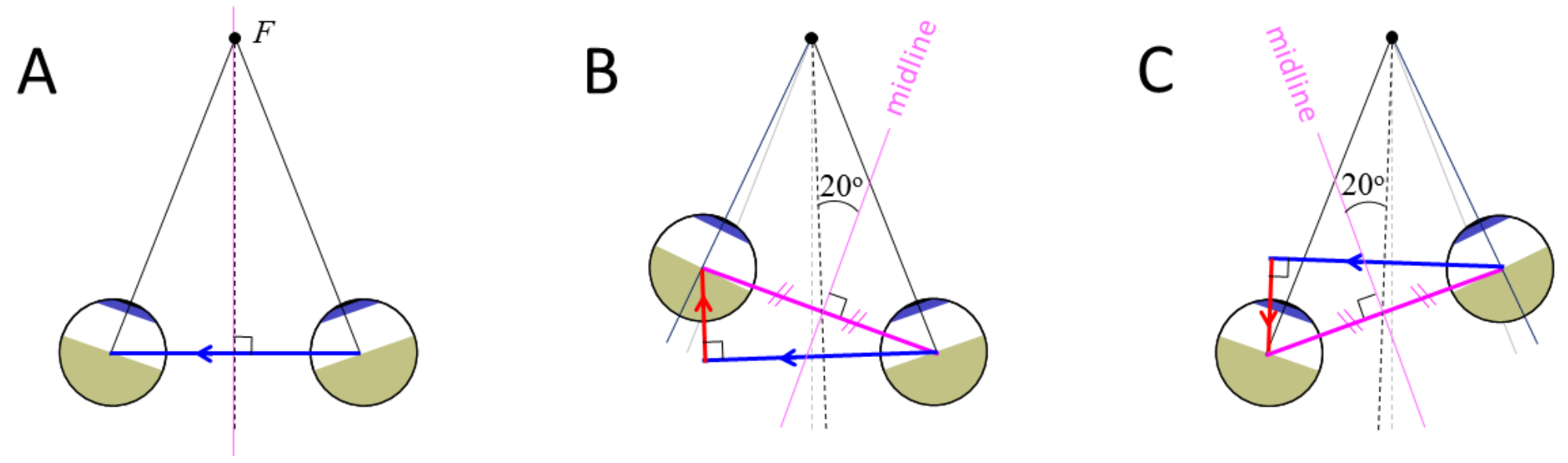

***Figure 4. Head rotation and ‘towards’ flow.*** *In binocular vision, ‘towards’ translation indicates fixation off the head-centric midline. (**A**): The observer binocularly fixates an object on the head-centric midline. The translation component is purely sideways. (**B, C**): Fixation 20º to the left (**B**) or right (**C**) of the midline. Now there is a ‘towards’ component expanding the image in the eye that is closer to fixation.*

Figure 5 is similar to Figure 3 but it zooms in on a small patch of the flow close to the fovea in order to illustrate in detail the behaviour of the ‘towards’ and ‘sideways’ components of flow. The illustration here is for points on a rough surface, i.e. one with random depth modulations within the surface, but the depth modulations are small compared to the mean distance to the surface. The top row (Figure 5A) shows flow that is generated when the surface fixated is 90 degrees from the epipole. Here, there is no 'towards' component but there are large 'sideways' and 'vergence' components. These two components give rise to optic flow or disparity along lines that can be considered parallel (given that we are considering a small patch). At the fixation point, these components cancel one another out exactly, so there is zero flow here. Away from the fixated point, the cancellation is not exact as shown in the first column (total flow). Flow is in one direction for points that are nearer than the fixation point and in the opposite direction for points that are further away. Of course, this is very familiar as a description of binocular disparities generated by a foveated surface, with positive values of 'horizontal' disparity indicating points that are further than the fixation point and negative values indicating nearer points.

However, whenever the surface is not exactly 90 degrees from the epipole (Figure 5B) there is a component of 'towards' flow and this means that flow (including disparity) is no longer 1-dimensional. The sum of the 'vergence' and 'sideways' components still carries useful information about the surface structure whereas the 'towards' component is predominantly useful as a source of information about the relative distance of the surface from the two optic centres. We discuss this division of labour in more detail in Section 3.1.

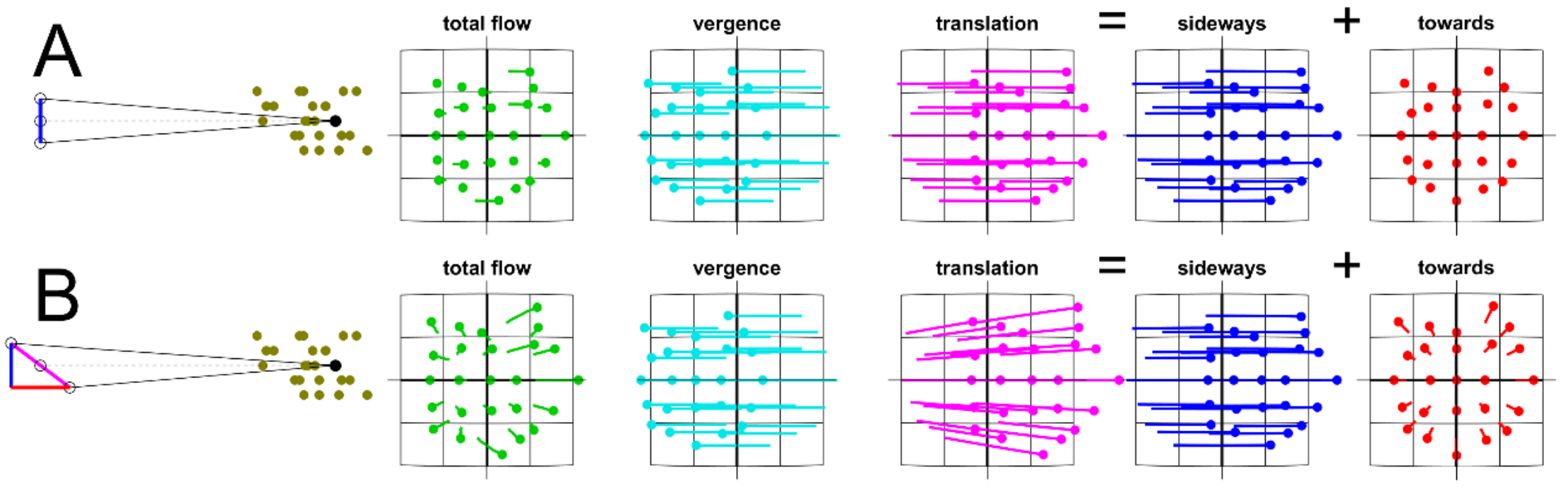

***Figure 5. Flow near the fovea****.* (**A**) *The fixation point is 90°from the epipole (i.e. on the headcentric midline for binocular viewing, hence there is no 'towards' component).* (**B**) *When fixation is at any other angle from the epipole there is a 'towards' component. The left-most column shows a top-down view of the eyes and a cloud of points around fixation. The 5 sets of axes in each row show different components of flow. In each case, lines link the projection of each point in Eye 1 with its projection in Eye 2 (the latter marked with a dot). The fovea is at the centre and grid lines mark ±5° and 10° from the fovea. Total flow (green) includes all flow components. Vergence flow (cyan) includes only rotational flow due to vergence (cyclovergence is zero in this example). Translational flow (pink) is made up of 'sideways' flow (blue) and 'towards' flow (red). The latter is zero in* (**A**)*.*

Finally, Figure 6 illustrates the same decomposition into 'vergence', 'towards' and 'sideways' flow but now applied to features across the whole retina. Away from the fovea, the 'sideways' component of flow is no longer a set of parallel lines, nor it is parallel to the vergence flow as it was in the foveal case (Figure 5). As a result, the directions of flow or disparity of points can vary over a wide range, as shown in Figure 6A (right hand column). The bottom row (Figure 6B) shows the consequences of fixating a point closer to the epipole, hence increasing the 'towards' component of the flow while reducing the 'sideways' component. This means that the total flow is close to the canonical 'towards' pattern of flow i.e. expansion outwards from the fovea (Figure 3C).

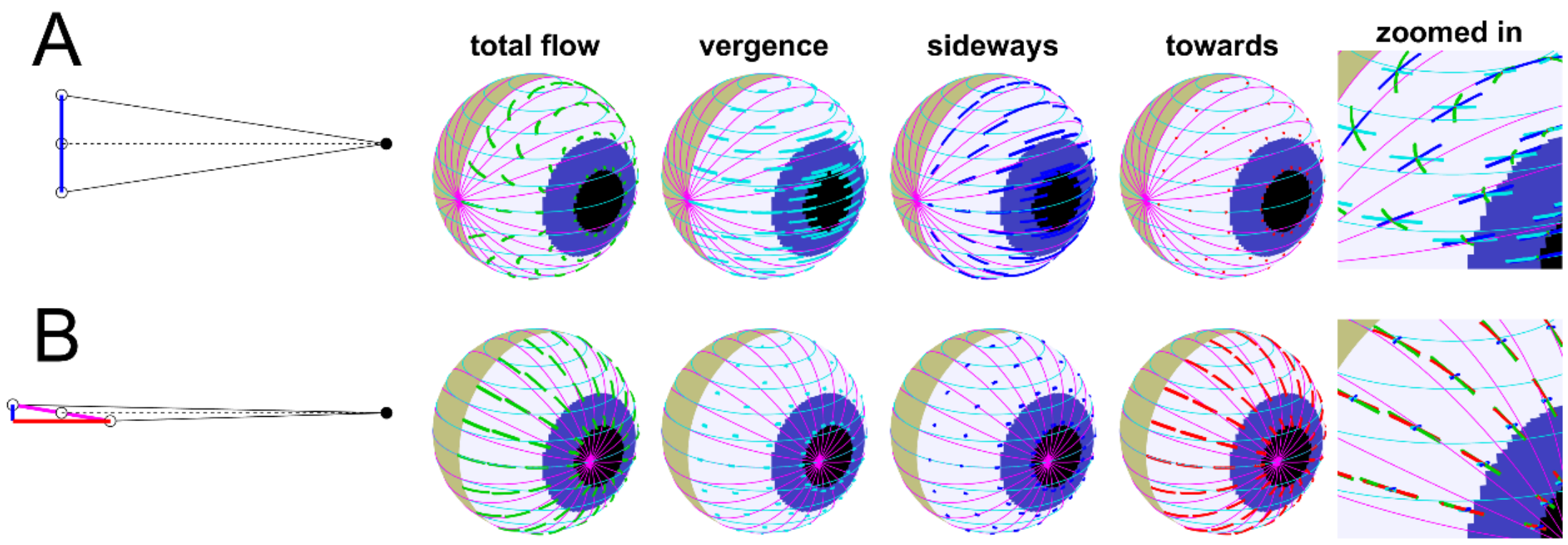

***Figure 6. Flow in the periphery.*** *As for Figure 5, but now showing flow across a whole hemisphere of the visual field. As before, the top row (***A***) shows flow for fixation 90 deg from the epipole ('straight ahead' in the binocular case) and the bottom row (***B***) shows fixation (much) closer to the epipole. Total flow is again shown in green, made up of components from vergence (cyan), sideways (blue) and towards (red) flow. The final plot in each row shows a zoomed-in view of part of the eyeball, showing how the green total flow is made up of these different components. On each eyeball, we have also marked the lines of latitude for vergence flow (light cyan circles) and the lines of longitude for translational flow for the specified epipole (light pink circles).*

## 2.2. In more detail

In this section, we describe 'sideways', 'towards' and 'vergence' flow in more detail and derive the statements that we have asserted above. Precise definitions and mathematical details are presented in the *Methods*. We also examine in detail the case of a small surface patch and how 'sideways' and 'towards' components of flow relate to surface slant in this case. Then, in the *Discussion*, we will compare the division of flow suggested here with other divisions of disparity or flow proposed in the literature. We will discuss whether there is a logic for choosing any one of these divisions over others, particularly when we consider flow in general, i.e. both binocular disparity and optic flow.

### 2.2.1. Coordinate systems and notation

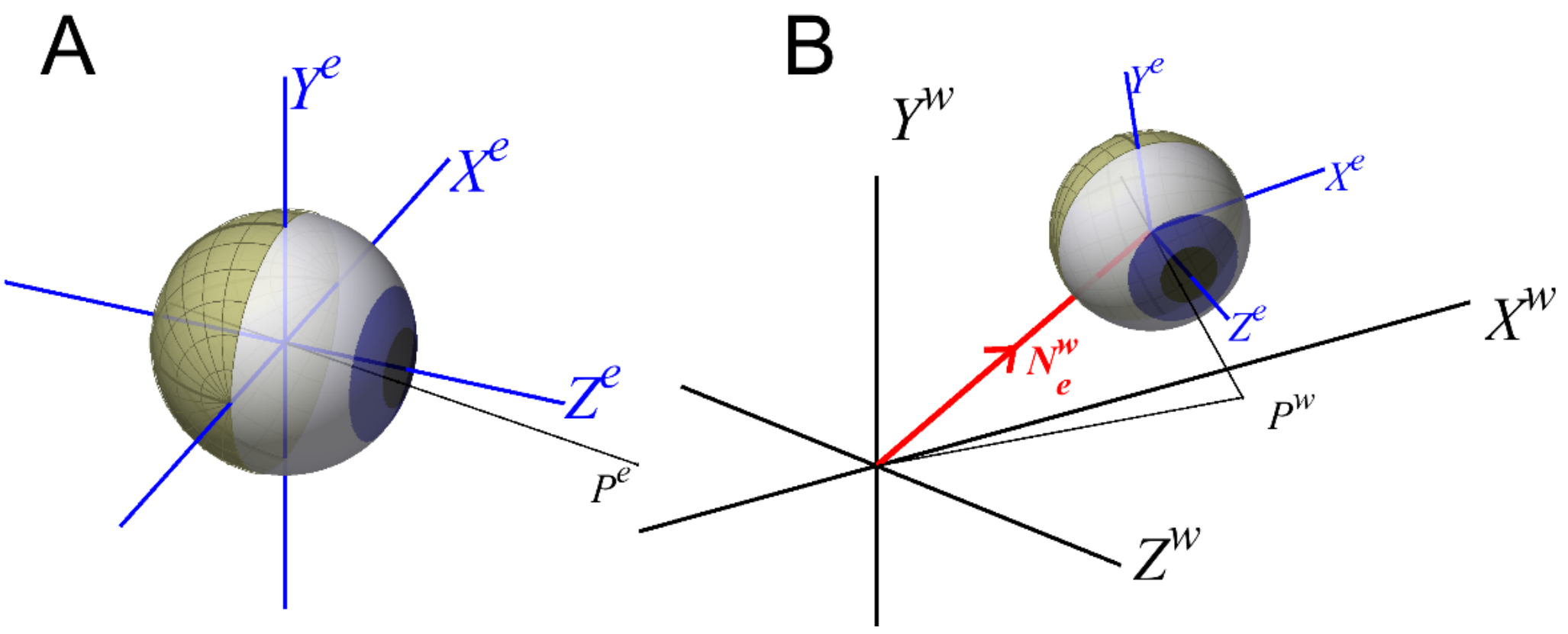

***Figure 7. Eye and world coordinate frames.*** *(**A**): An eye-centered Cartesian coordinate system ($X^e$, $Y^e$, $Z^e$). $Z^e$ is the visual axis; $X^e$ is 'horizontal' on the retina, and $Y^e$ is 'vertical'. (**B**): world-centered Cartesian system ($X^w$, $Y^w$, $Z^w$), related to ($X^e$, $Y^e$, $Z^e$) by a translation, $N_e^w$, and a rotation (Eq 1). P is an example scene point. To help make clear the orientation of the eyeball, we have marked the pupil and iris on the front, and drawn an azimuth-latitude/elevation-longitude angular coordinate system on the retina (Read et al 2009).*

We use superscripts to indicate the coordinate system. Figure 7A shows an eye-based coordinate frame ($X^e$, $Y^e$, $Z^e$). By definition, the origin of this system is the nodal point of the eye. The $Z^e$-axis is the visual axis, which in this idealised eye runs from the fovea through the nodal point and then out through the centre of the pupil; $X^e$ and $Y^e$ define the horizontal and vertical meridians on the retina. Figure 7B shows this eye within a world-based coordinate frame ($X^w$, $Y^w$, $Z^w$). A point P can be defined in either coordinate frame ($\mathbf{P}^e$ or $\mathbf{P}^w$). The eye-centered coordinates of a point, $\mathbf{P}^e$, specify the projection of that point onto that eye. Note that the location in the eye to which the point projects is defined solely by the direction of the vector $\mathbf{P}^e$, not by its length.

The relationship between the world-centered coordinates $\mathbf{P}^w$ of the point P and its eye-centered coordinates, $\mathbf{P}^e$, is:

$$\mathbf{P}^e = R_w^{\ e} (\mathbf{P}^w - \mathbf{N}_e^{\ w}) \qquad \text{Eq 1}$$

or conversely:

$$\mathbf{P}^w = R_e^{\ w} \mathbf{P}^e + \mathbf{N}_e^{\ w}$$

where the vector $\mathbf{N}_e^{\ w}$ specifies the location of the eye's nodal point in world-centered coordinates. The matrix $R_e^{\ w}$ is the rotation matrix specifying the eye's orientation in world-centered coordinates, and $R_w^{\ e}$ is its inverse.

For most of this paper, we will work in the coordinate frame of the cyclopean eye, whose location we define to be exactly halfway between Eye 1 and Eye 2 (Figure 2A). We define the visual axis of the cyclopean eye to point at the fixation point (or, more generally, the *pseudofixation point*, the point midway between the visual axes at their point of closest approach, for non-fixating eye postures; see *Methods* for details). This fixation constraint specifies two of the three degrees of freedom for the cyclopean eye orientation. Finally, we define the cyclopean eye as having zero torsion in its own coordinate system. This means that if the two eyes have any cyclovergence, they have equal and opposite cyclotorsion in cyclopean coordinates. Note that the cyclopean eye may still have non-zero torsion in world-centered or (more pertinently for binocular vision) head-centered coordinates.

We will consider a scene point that projects to location $\mathbf{P}^c$ in the cyclopean eye. To find the flow of this point, we need to find its projection in Eye 1 and Eye 2. The flowlines move from the projection $\mathbf{P}^1$ in Eye 1, through the projection $\mathbf{P}^c$ in the cyclopean eye, to the projection $\mathbf{P}^2$ in Eye 2. Assuming a static world, as we do in this paper, flow is generated by the translation and rotation of the eyes. In the cyclopean frame, Eye 2 is at the epipolar vector, $\mathbf{N}_2^c = \boldsymbol{\varepsilon}^c$, while Eye 1 is at $\mathbf{N}_1^c = -\boldsymbol{\varepsilon}^c$. The rotation matrices $R_1^c$ and $R_2^c$ describe the orientation of each eye in the cyclopean frame.

In order to plot the flowlines, we can imagine that the cyclopean eye stays constant and the world moves around it. The coordinate system of Eye 2 is related to cyclopean coordinates by a rotation $R_2^c$ and a translation $+\boldsymbol{\varepsilon}^c$. Thus, to find the projection of $P^c$ into Eye 2, we apply a translation of $-\boldsymbol{\varepsilon}^c$ and the inverse rotation $R_c^1$ (Figure 8). Thus, a scene point with cyclopean coordinates $\mathbf{P}^c$ projects into Eyes 1 and 2 as follows:

$$\boldsymbol{P}^1 = R_c^1 (\boldsymbol{P}^c + \boldsymbol{\varepsilon}^c); \qquad \boldsymbol{P}^2 = R_c^2 (\boldsymbol{P}^c - \boldsymbol{\varepsilon}^c) \qquad \text{Eq 2}$$

This is consistent with Eq 1, replacing *e* with 1,2 for the two eyes, world coordinates *w* with cyclopean coordinates, *c*, and setting $\mathbf{N}_1^c = -\boldsymbol{\varepsilon}^c$, $\mathbf{N}_2^c = \boldsymbol{\varepsilon}^c$.

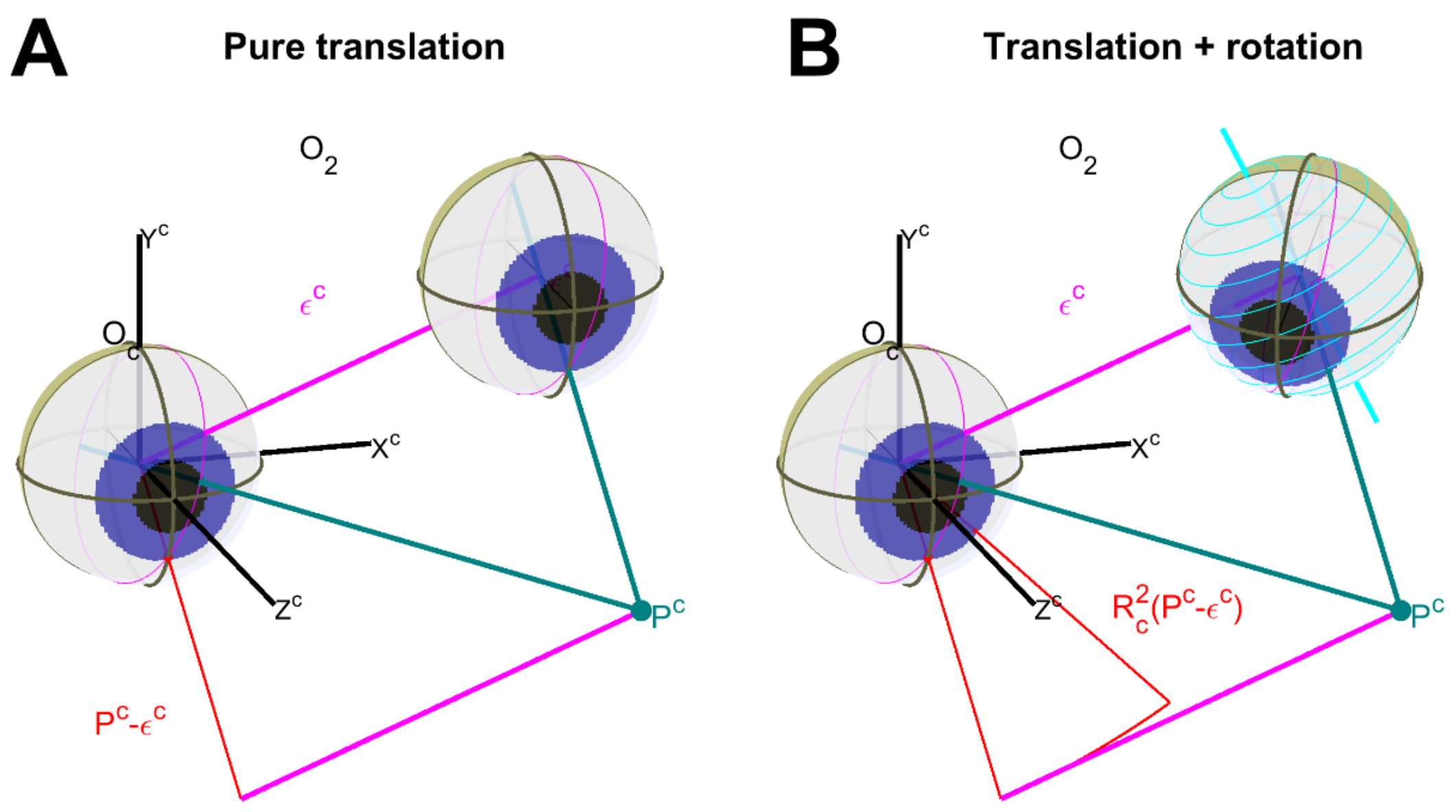

***Figure 8. Relating the projection of a scene point, P, in the cyclopean eye and Eye 2.*** *Eye 2 is at $O_2$, offset from the cyclopean eye, at $O_c$, by the vector* $\boldsymbol{\varepsilon}$*. In* (**A**)*, the eyes have the same orientation, so the scene point* $\mathbf{P}^c$ *projects to Eye 2 at the same location as the point* $(\mathbf{P}^c\text{-}\boldsymbol{\varepsilon})$*, shown in red, projects to the cyclopean eye. In* (**B**)*, Eye 2 is also rotated relative to the cyclopean eye as described by the rotation matrix* $R_2^c$*. Hence, the inverse rotation* $R_c^2$ *would have to be applied to the vector* $(\mathbf{P}^c\text{-}\boldsymbol{\varepsilon})$ *if it were to project to the same location in the cyclopean eye as* $P^c$ *projects to in Eye 2. Therefore, the vector* $\mathrm{R}_c^2(\mathbf{P}^c\text{-}\boldsymbol{\varepsilon})$ *describes where the scene point* $P^c$ *projects to in Eye 2. Similarly,* $\mathrm{R}_c^1(\mathbf{P}^c\text{+}\boldsymbol{\varepsilon})$ *would describe where the scene point* $P^c$ *projects to in Eye 1. The 'flow' generated by the rotation and translation of the eye is the curve joining these two projections.*

In the next section, we will decompose the flow described by Eq 1 into 4 components. First, we give a more formal account of how to decompose translation into sideways and towards components, as described above. Second, we describe decomposing rotation into vergence and cyclovergence components.

### 2.2.2. Decomposing translation into sideways+towards components

Pure translational flow is obtained by ignoring the rotational component. From Eq 1 with no rotation, we see that flow in this case is:

$$\boldsymbol{P}^{1\mathrm{Tr}} = \boldsymbol{P}^{c} + \boldsymbol{\varepsilon}^{c} \quad ; \quad \boldsymbol{P}^{2\mathrm{Tr}} = \boldsymbol{P}^{c} - \boldsymbol{\varepsilon}^{c} \qquad \text{Eq 3}$$

As is clear from Figure 8A, the flowlines here are along a great circle through the image point in question, $\mathbf{P}^c$, and the epipole, $\boldsymbol{\varepsilon}^c$. Some examples are shown below in Figure 10 (pink circles).

In Figure 3, we introduced the idea of dividing the translational component of flow into ‘sideways’ and ‘towards’ components that are caused by orthogonal components of translation (Figure 3A):

$$\boldsymbol{\varepsilon} = \boldsymbol{s} + \boldsymbol{t}$$

$\boldsymbol{\varepsilon}$ is the translation between the cyclopean eye and Eye 1 or Eye 2; **t** is the component of translation towards the fixated point, defined as being parallel to the visual axis of the cylopean eye; **s** is the component of translation orthogonal to **t**. In cyclopean coordinates, $\mathbf{t}^c$ lies along $Z^c$ while $\mathbf{s}^c$ is in the $X^cY^c$ plane. Put another way, for pure ‘towards’ flow, the epipole is at the cyclopean fovea (Figure 3C). For pure ‘sideways’ flow, the epipole is at an eccentricity of 90° on the cyclopean retina, i.e. on the white/gold boundary in our figures (Figure 3D). In general, as we have seen, translational flow will be a mixture of 'towards' flow and 'sideways' flow (Figure 3B).

From Eq 3, pure towards flow is:

$$\boldsymbol{P}^{1\mathrm{To}} = \boldsymbol{P}^{c} + \boldsymbol{t}^{c};\ \boldsymbol{P}^{2\mathrm{To}} = \boldsymbol{P}^{c} - \boldsymbol{t}^{c}$$

Since by definition $\boldsymbol{t}$ points along the cyclopean visual axis, these flowlines are along a great circle through $\mathbf{P}^c$, and the cyclopean fovea. This is shown in red in Figure 10. Similarly, pure 'sideways' flow is:

$$\boldsymbol{P}^{1\mathrm{Si}} = \boldsymbol{P}^{c} + \boldsymbol{s}^{c};\ \boldsymbol{P}^{2\mathrm{Si}} = \boldsymbol{P}^{c} - \boldsymbol{s}^{c}$$

To find where this is on the retina, we draw a great circle through the fovea and the epipole, and see where this intersects the 90°-eccentricity meridian. This is the vector $\mathbf{s}^c$. Sideways flow is along the great circle through $\mathbf{P}^c$ and $\mathbf{s}^c$, as shown in blue in Figure 10.

### 2.2.3. Decomposing rotation into vergence+cyclovergence components

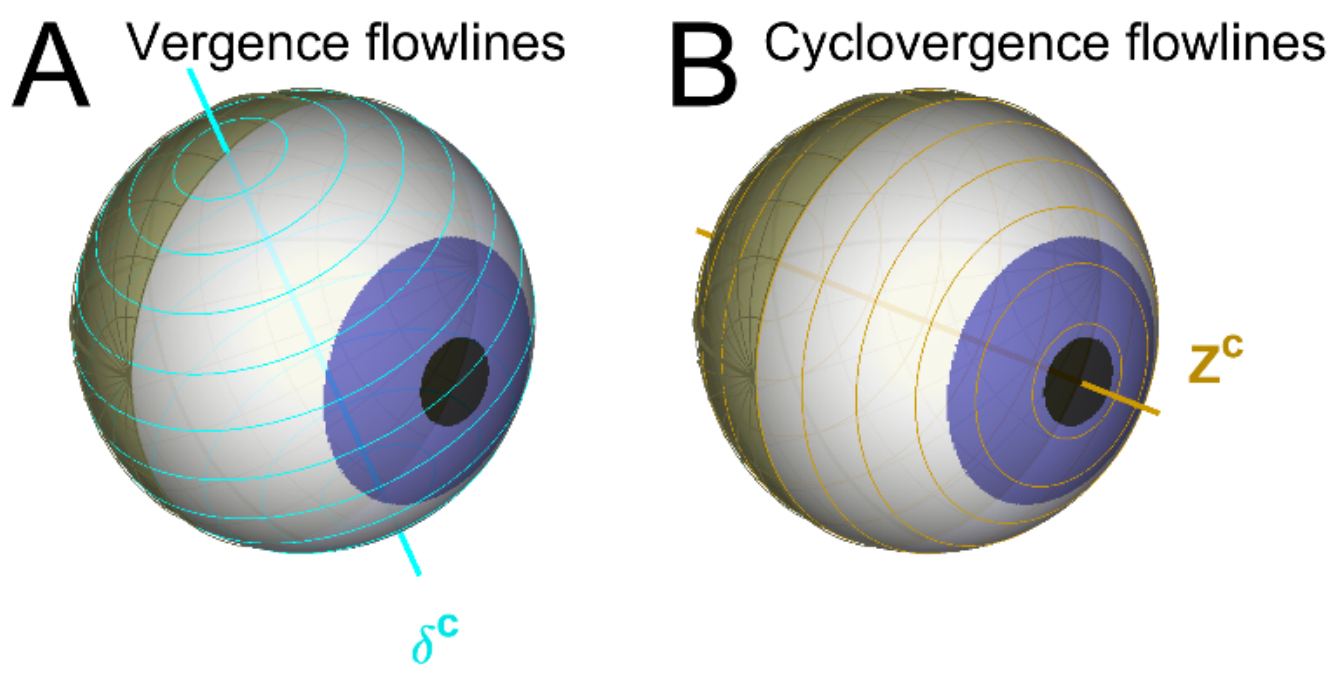

***Figure 9. Decomposing rotational flow into vergence and cyclovergence components.*** *(**A**) For a fixating observer, the axis for vergence ($\delta^c$) is at 90º from $\boldsymbol{Z}^c$ . (**B**)For cyclovergence, the axis is $\boldsymbol{Z}^c$.*

From Eq 1 with no translation, we see that pure rotational flow is:

$$\boldsymbol{P}^{1\mathrm{Ro}} = \mathrm{R}_c^{\,1}\, \boldsymbol{P}^c \qquad \text{Eq 4}$$
$$\boldsymbol{P}^{2\mathrm{Ro}} = \mathrm{R}_c^{\,2}\, \boldsymbol{P}^c$$

In Figure 2, we considered the case of a fixating eye with zero cyclovergence. The rotational component of flow was then about an axis orthogonal to the plane of regard. In general, e.g.

including non-biological eye movements, this may not be true. However, it is still helpful to decompose the rotational component of flow into vergence and cyclovergence components. We define the vergence axis $\boldsymbol{\delta}$ to be orthogonal to the plane of regard defined by the visual axes of the two eyes (plane $O_1O_2F$ in Figure 2A; see *Methods* and Figure A1 for more details). Since the cyclopean eye is defined such that its visual axis, $\boldsymbol{Z_c}$, lies in this plane of regard, this means that the vergence axis must be orthogonal to the cyclopean visual axis, at 90° eccentricity in the cyclopean eye (Figure 9A):

$$\boldsymbol{\delta}.\boldsymbol{Z_c} = 0$$

The axis for cyclovergence rotations is the cyclopean visual axis, $\boldsymbol{Z_c}$ (Figure 9B). As we show in the *Appendix* (Eq A6 - Eq A8), the rotation matrices for the two eyes can be decomposed into vergence and cyclovergence components:

$$R_c^1 = T\, V_c^1\,; \quad R_c^2 = T^T\, V_c^2$$

Thus, the rotational component of flow can be further decomposed into vergence flow:

$$\boldsymbol{P}^{1Ve} = V_c^1\, \boldsymbol{P}^c \quad ; \quad \boldsymbol{P}^{2Ve} = V_c^2\, \boldsymbol{P}^c \qquad \text{Eq 5}$$

and cyclovergence flow:

$$\boldsymbol{P}^{1Cy} = T\, \boldsymbol{P}^c \quad ; \quad \boldsymbol{P}^{2Cy} = T^T\, \boldsymbol{P}^c$$

i.e. when there is non-zero cyclovergence, the torsion of Eye 1 and Eye 2 is equal and opposite with respect to the cyclopean eye.

Figure 9 shows flowlines for vergence (cyan) and cyclovergence (gold) components. Note that we do not show the direction of the net rotational flow. This is because the net rotation is, in general, different in the two eyes. Because we chose to define the cyclopean eye to point to the pseudofixation point, the rotation angle from the cyclopean eye to Eye 1 is in general different from that to Eye 2. This is apparent in the example drawn in Figure 2A. This makes vergence different from all the other components, where we have defined the cyclopean eye to be exactly midway between Eye 1 and Eye 2. Because of this symmetry, cyclovergence flow, and sideways and towards flow (and therefore also net translational flow) all have the same magnitude and opposite direction for the two eyes. However, vergence flow is not necessarily equal in magnitude in the two eyes. This means that if there is any cyclovergence, the net rotational flow resulting from vergence+cyclovergence will be about a different axis in the two eyes. Thus, there is no single direction of "total rotational flow" for us to show. Full details are given in the *Methods* (Figure A2).

Like other flow components, we have chosen to define cyclovergence relative to the cyclopean eye. We are only concerned here with flow, i.e. flow caused by *changes* in torsional state between the cyclopean eye, Eye 1 and Eye2 so issues relating to the torsion of the eye relative to the head in different eye positions do not concern us here (Tweed, 1997; Schreiber et al, 2001; Banks et al, 2015).

Thus, in general, flowlines at a given point in the visual field are made up of contributions from four components, two translational components, 'sideways' and 'towards', along great circles on the retina, and two rotational components, 'vergence' and 'cyclovergence', along circles of latitude.

### 2.3. Biological constraints: The consequences of fixation

In this section, we introduce constraints that apply to most animals, as we discussed in the Section 2.1. We now treat these constraints more formally and consider differences between optic flow and binocular viewing.

In the previous section, we considered general flow, with the eye free to move with 6 degrees of freedom. The vast majority of animals, from insects through to humans, do not do this; instead they stabilise their gaze on an object as they move, followed by a very rapid saccade. Binocular fixation on a point is one example of this yoked pattern of eye translation and eye rotation.

If the eyes are fixating on a single point in space, then the epipolar vector must lie in the same plane as the visual axes, the plane of regard (see *Methods*, Figure A1). Since the vergence axis, $\boldsymbol{\delta}$, is defined to be orthogonal to this plane, this means that the vergence axis is also now orthogonal to the epipolar vector, as well as to the cyclopean visual axis:

$$\boldsymbol{\delta}.\boldsymbol{\varepsilon} = 0 \text{ for fixation.}$$

Relating this to the 'sideways' and 'towards' directions, for fixation, if we write:

$$\boldsymbol{\varepsilon}^c = (s_x, s_y, t)$$

then

$$\boldsymbol{\delta}^c = (s_y, -s_x, 0).$$

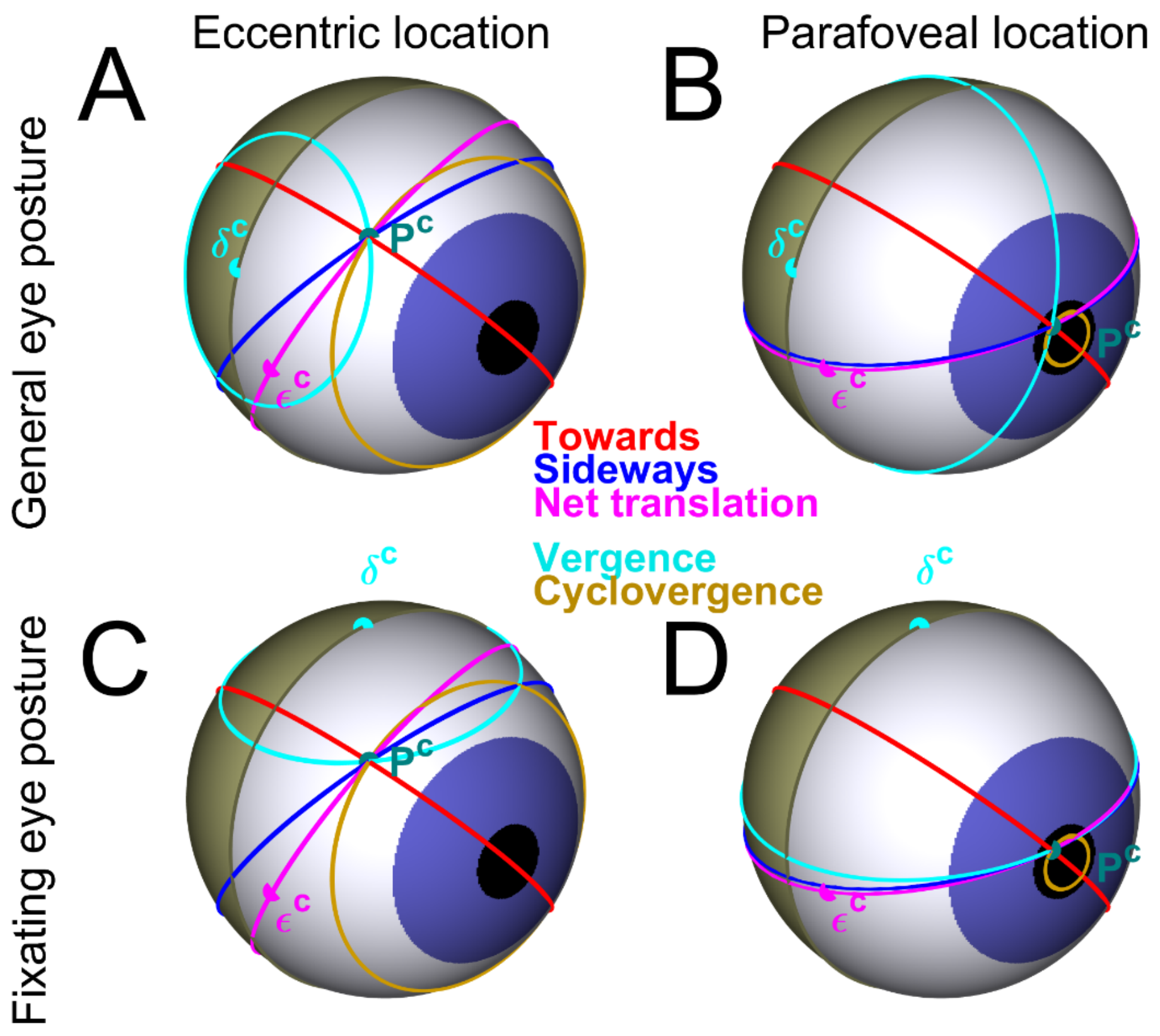

***Figure 10. The components of flow for two different eye postures and image locations****. Top row (***A,B***): A general eye posture (not fixating). Bottom row (***C,D***): A fixating eye posture. Left column (***A,C***): eccentric image location. Right column (***B,D***): parafoveal image location. The coloured circles show the direction of flow components at the scene point* $\boldsymbol{P}^c$.

*Pink lines show net translational flow which is the sum of 'towards' flow (red) and 'sideways' flow (blue). $\varepsilon_c$ is the epipole for translational flow. Cyan lines show vergence flow, gold lines show cyclovergence flow which together make up rotational flow (see Figure 2). For the fixating eye, the vergence axis ($\delta_c$) is orthogonal to the epipolar vector. Thus, near the fovea, vergence flow is roughly parallel to net translational flow (***D***). The 'sideways' flow epipole is always on the gold/white border, 90$^o$ from the fovea. The 'towards' flow epipole is at the fovea.*

In the case of fixation with pure 'towards' motion, $\delta^c$ =0, meaning the vergence axis is undefined. This is because if a fixating eye is translating directly along the line of sight, there can be no vergence movements, since any such rotation would move the fovea off the fixated object. For a general translation consisting of both 'sideways' and 'towards' components, the vergence axis, $\boldsymbol{\delta}^c$, *will* be defined and the angle of rotation about $\boldsymbol{\delta}^c$ will vary depending on the distance to the fixation point. When the fixation point is at infinity, the amplitude of the rotational motion becomes zero. The direction of flow in this case is along the translational lines of longitude, i.e. in the same direction as epipolar lines.

Figure 10 illustrates the consequences of applying the fixation constraint as the eye moves. In each panel, we have marked the direction of flow components at a particular image point $\mathbf{P}^c$. The left column is for an eccentric image point, $\mathbf{P}^c$, whereas the right column is for a point near the fovea. The top row shows flow when the eye translates and rotates in a general way (6 degrees of freedom) whereas the bottom row show flow when the eye fixates a point as it translates. The clearest consequence of fixation is near the fovea (Figure 10D) where the direction of translational flow (pink), 'sideways' (blue) and 'vergence' flow (cyan) are all approximately parallel at the point $\mathbf{P}^c$ (Figure 10B) whereas, for a general rotation and translation from Eye 1 to Eye 2, ‘vergence’ flow is in quite a different direction.

In Section 3.2, we explore reasons why, for an animal that is moving in relation to a fixation point and making judgements about the depth of points relative to the fixation point, the division of flow described here may have some practical advantages in relation to the control of movement. But first, in the next section, we consider how this division relates to heuristics that have been proposed to recover information about the slant of a small surface patch viewed binocularly or by a moving observer.

### 2.4. Flow for a small surface patch

In the field of binocular vision, many previous workers have pointed out that information about the slant of a surface can be derived from information about its ‘horizontal’ and ‘vertical’ size ratios (Gillam and Lawergren, 1983; Rogers and Bradshaw, 1993; Backus et al, 1999; Kaneko and Howard, 1996). Now that we are considering optic flow in a moving observer, some of the approximations that have been used in the binocular case do not apply, so it is worth revisiting this issue and being clear about the information that is present in the flow field in relation to surface slant. We will see that the important coordinates can be described in terms of the epipolar geometry alone, i.e. they are determined by the location of the optic centres in the scene and not the orientation of the eyes. The directions specified by the relevant coordinate frame are ‘epipolar’ (in the epipolar plane so, for a fixated object, in the plane of regard) and ‘ortho-epipolar’ (perpendicular to this). To simplify the description, we consider a viewer fixating a point on a surface. This allows us to refer to ‘towards’ and 'sideways' flow, which we have defined in relation to a fixated point, but the geometry we describe and Eqn 6 do not depend on the orientation of the eyes.

The surface is small so it can be considered to be locally planar. We assume zero cyclovergence, so that flow has a 'towards' component, a 'sideways' component and a 'vergence' component which ensures that the point remains fixated (similar to 'affine' divisions of flow in a small region of the visual field, Koenderink and van Doorn (1976, 1991); Shapiro, Zisserman and Brady (1995)). At the fovea, as we have seen, sideways and vergence flowlines are parallel (e.g. Figure 10D). Now consider 4 image-points arranged on a cross around the fovea (Figure 11). Two points shown in pink in Figure 11B, are offset relative to the centre of the cross in a direction parallel to the epipolar vector, $\boldsymbol{\varepsilon}^c$, hence we will call them 'E-points'. These points lie along the sideways/vergence flowline through the fovea. The scene-points corresponding to the E-points in the image lie in the plane of regard. The cyan points in Figure 11 are offset in the orthogonal direction and so we call those 'O'-points. On the retina, this direction is orthogonal to the plane of regard and hence parallel to the vergence vector, $\boldsymbol{\delta}$.

We define the 'E-size' and 'O-size' of the cross as the angular separation between the E-points and O-points respectively. In the cyclopean image, by construction, the cross's arms are of equal length, so E-size = O-size, but we now consider what happens in the two eyes' images of the scene points corresponding to the 4 points of this cross. We will consider the ratio of the sizes in each eye, in other words the 'E-size ratio' (ESR) and the 'O-size-ratio' (OSR).

The vergence flow simply slides all points by the same amount along the sideways/vergence flowlines, shown in blue in Fig 11. The displacement will be in opposite directions in the two eyes. Thus, vergence flow changes neither E-size nor O-size; it simply re-positions the image on the retina, re-centering it on the fovea after it has been displaced by sideways flow.

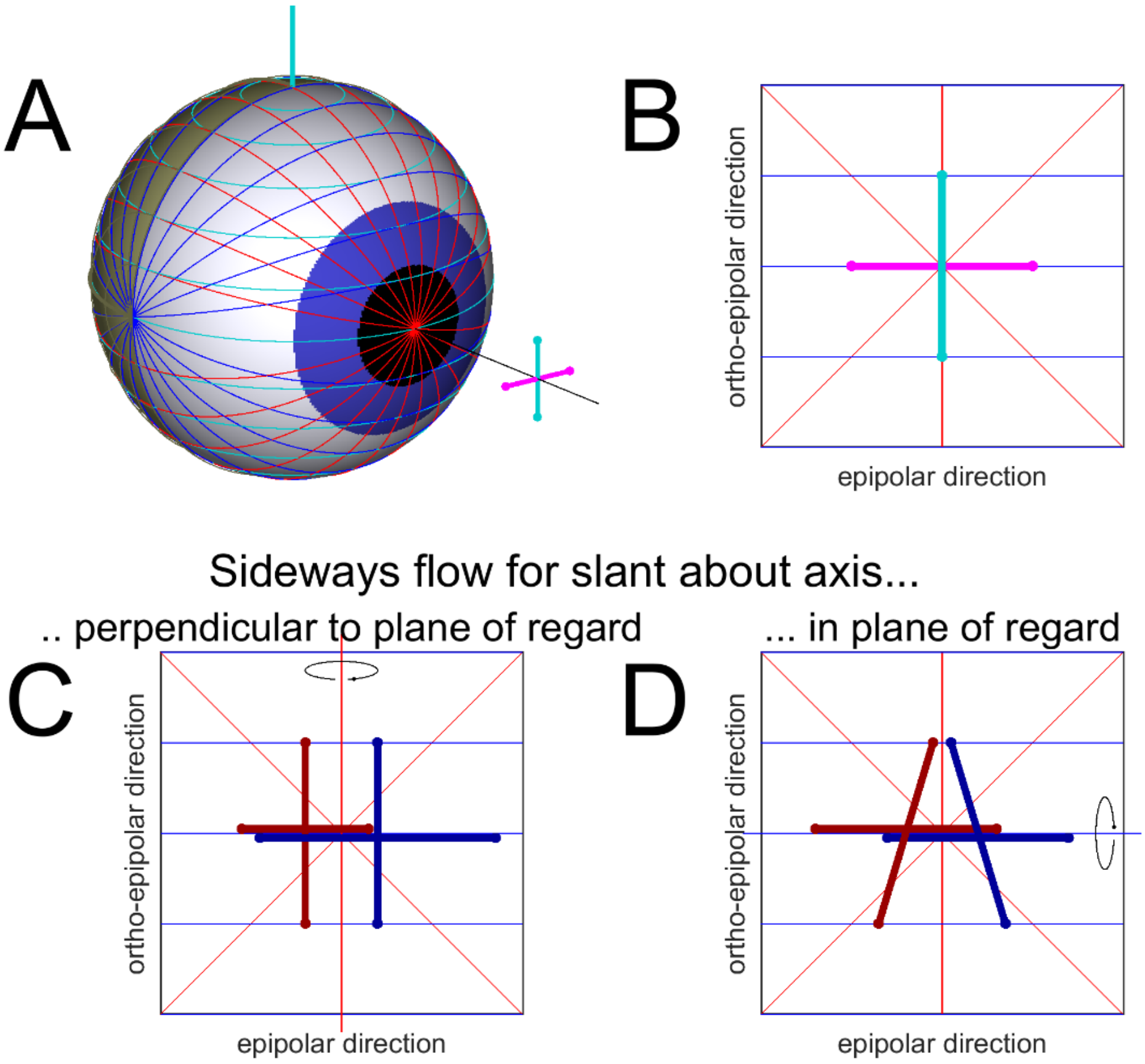

***Figure 11. Sideways flow for a small patch****. (***A***): Cyclopean eye showing directions of towards (red), sideways (blue) and vergence (cyan) flow. Four points are shown arranged on a cross about fixation (size exaggerated for clarity). (***B,C,D***): Projections of these points on the retina.* **B***: Cyclopean image locations for no surface slant.* **CD***: How sideways flow shifts points away from the cyclopean images, in Eye 1 (red) and Eye 2 (blue), when the cross is slanted about an axis, (***C***), perpendicular to the plane of regard or, (***D***), in the plane of regard. In (***C***), the left-hand end of the cross is further from the viewer so, considering*

*'sideways' flow alone as shown here, the images in the two eyes lie closer together than for the right-hand end. The effect is to compress the image in Eye 1 and expand it in Eye 2 along the epipolar axis, with no change along the orthogonal axis. In* **D***, the top of the cross is further than the bottom; this shears the images without compression or expansion. Vergence flow would then shift the images equally and opposite in both eyes, such that the center of the cross moves back to the fovea. Towards flow would shift image-points along the red, radial flowlines. In* **C** *and* **D***, the images for the left and right eyes have been given a slight vertical shift for clarity.*

'Towards' flow shifts image-points radially along the red flowlines. This expands the image in one eye and contracts it in the other. The expansion/contraction is isotropic, so if there is *only* 'towards' flow, the E-size ratio and O-size ratio will be equal to each other: ESR = OSR. The size ratio is simply due to the relative distance of the patch to the two eyes, i.e. ESR = OSR = (size in Eye 1) / (size in Eye 2) = (distance from P to Eye 2) / (distance from P to Eye 1).

For 'sideways flow', a crucial difference from vergence flow is that the magnitude of the displacement for each image-point will depend on the cyclopean distance to that image point. Thus, when the two E-ends of the cross are at the same cyclopean distance, the displacement will be the same and there will be no change in E-size in either eye. However, when the E-points are at different cyclopean distances, the E-size will be expanded in one eye and contracted in the other (Figure 11C). Thus, the E-size ratio will be different from 1. This occurs when the surface is slanted about an axis perpendicular to the plane of regard. Slant about any axis lying *in* the plane of regard cannot change the distance of the E-points, and so will not alter the E-size. It will, of course, change the distance of the O-points, causing the image of the cross to shear in opposite directions in the two eyes (Figure 11D). The resulting disparity gradient is a retinal cue to the surface slant about an axis in the plane of regard.

Putting these elements together, if the fixated surface is face-on to the cyclopean gaze *or* slanted only about an axis lying in the plane of regard, the E-size ratio will equal the O-size ratio. Slant about an axis orthogonal to the plane of regard will alter the E-size ratio (because of the unequal sideways-flow) but not the O-size ratio. Thus, "ESR:OSR", the ratio of E-size ratio to O-size ratio, contains information about slant about an axis orthogonal to the plane of regard (the vergence axis, in our notation). If ESR=OSR, then the surface is not slanted about an axis orthogonal to the plane of regard (although it may be slanted about an axis lying in this plane). OSR on its own carries information about the amount of towards flow (relative to fixation distance). If OSR=1, then there is no 'towards' flow, i.e. the observer is moving past the object without approaching it. The only situation where ESR and OSR fail to give any information about surface orientation are when an observer is approaching the fixated object directly. Then, the plane of regard is undefined and ESR=OSR regardless of surface slant.

When the surface is *not* facing the cyclopean point, the E-size in the two eyes will be different. The difference indicates the direction of slant and, in one interesting case, the magnitude. When the surface is parallel to the epipolar vector, $\varepsilon^c$, OSR and ESR have a simple relationship: ESR = OSR$^2$. Essentially, two effects combine to influence ESR and for this particular slant the effects are of the same magnitude. One effect is caused by the ratio of distances to the two eyes, as we have seen, while the other occurs because a slanted surface is foreshortened by different amounts in the two eyes. Figure 12 illustrates this.

Figure 12A shows the cyclopean eye along with the direction of the epipole, $\varepsilon^c$, and a scene point, $P^c$. Both the epipole and the scene point are chosen to be arbitrary, with no special relationships to each other or to the fovea. The great circle through the epipole and the scene point is shown in teal; this gives the epipolar direction through $P^c$. This same great circle is shown in the plane of the page in Figure 12B. In Figure 12C and D, we zoom in on the patch in question. Figure 12C shows the situation where the surface patch lies at right angles to the line of sight and hence the surface patch is tangent to the great circle shown in Figure 12B. To calculate the E/OSR in this case, we note that although the patch is shown large for illustration, it is really intended to be infinitesimal, so the angle it subtends in each eye is inversely proportional to its distance from that eye, $d_e$. Therefore, in this situation, ESR=OSR= $d_2/d_1 = \sin\zeta_1/\sin\zeta_2$. Even when the patch is not fixated, the same geometry applies.

Figure 12D shows how the situation changes when the patch rotates about an axis orthogonal to the plane of regard, so that it is now parallel to the epipolar vector. The patch is now foreshortened in each eye. Its effective size in Eye 1, measured along the epipolar direction (i.e. in the plane of regard), is shown by the red line. Because the patch is infinitesimal, the lines of sight from $O_1$ to either end of the patch are effectively parallel, and therefore all three shaded angles marked in Figure 12D are approximately equal to $\zeta_1$. Foreshortening therefore reduces the effective E- size of the patch in Eye 1 by a factor of $\sin\zeta_1$ and in Eye 2 by a factor of $\sin\zeta_2$. The ESR therefore acquires an *additional* factor of $\sin\zeta_1/\sin\zeta_2$. Thus, for any patch slanted so as to be parallel with the epipolar direction:

$$\mathrm{ESR}=\mathrm{OSR}^2. \qquad \text{Eqn 6}$$

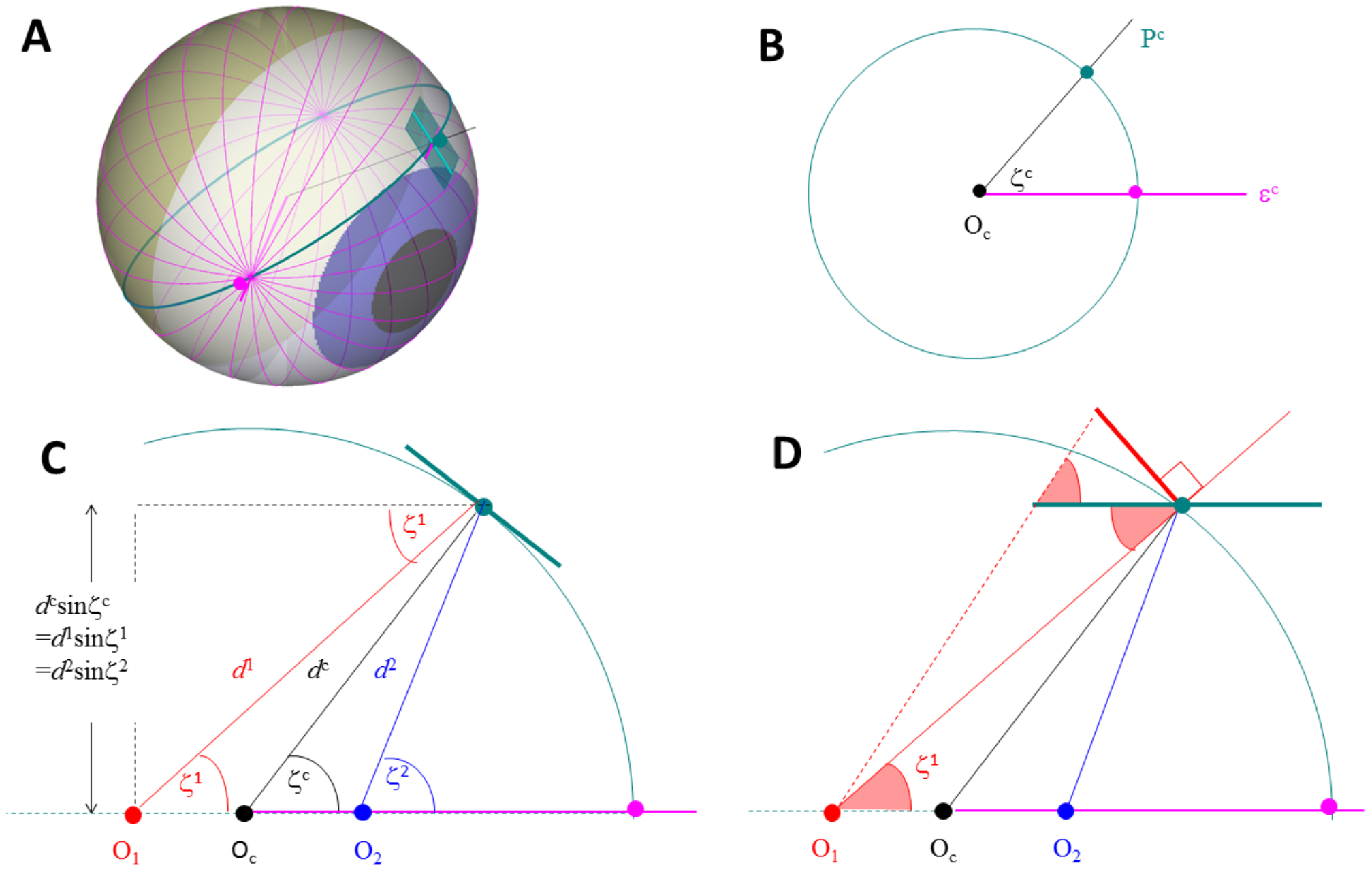

***Figure 12. The effect of surface slant on epipolar size ratio (ESR).*** *(A) shows the cyclopean eye, the epipolar planes (pink) and a surface patch and the epipolar plane through it shown in teal. That epipolar plane is shown in* ***(B)*** *including the angle between the epipolar vector, $\varepsilon_c$, and the direction of the patch, $P_c$. The distance to the patch differs between Eye 1 and Eye 2, as shown in* ***(C)****, where $d_2/d_1 = \sin\zeta_1/\sin\zeta_2$. When the patch is facing the cyclopean point, as shown here, $d_2/d_1$ is the only factor that affects ESR. (D) When the patch is slanted so that it is parallel to the epipolar vector, $\varepsilon_c$, there is an additional foreshortening factor as illustrated by the red line. The red line is foreshortened by $\sin\zeta$ relative to the slanted surface. So, the ratio of foreshortening effects in the two eyes is $\sin\zeta_1/\sin\zeta_2$, just like the distance effect in (C). See text for details.*

The literature on this relationship can be quite confusing. It is often claimed that a small fronto-parallel patch obeys the rule that its ‘horizontal’ size ratio (HSR) and ‘vertical’ size ratio (VSR) are related in the same way as Eqn 6, i.e. that HSR = VSR$^2$. However, if ‘horizontal’ and ‘vertical’ refer to a retinal coordinate system (e.g. Howard and Rogers, 1995) then it is not always the case that HSR is the same as ESR nor that VSR is the same as OSR and, hence, it is not true in general that HSR = VSR$^2$ for frontoparallel surfaces. We illustrate this point by considering a very extreme case where the eyes fixate a patch close to the epipole (something that is quite natural for optic flow but impossible in binocular vision as it would mean looking along the interocular axis). In this extreme case, ‘horizontal’ can be almost orthogonal to the epipolar direction so the difference between ESR and HSR is dramatic. Figure 13 shows how, near the epipole, the direction of epipolar planes changes very rapidly over small retinal distances. This exposes very clearly any differences between retinal and epipolar coordinate frames. The situation shown in Figure 13 corresponds to a monocular observer walking towards a point and fixating on it. As they walk, his or her head bobs up and down and from side to side, adding small lateral components to a translation vector whose main component is towards the fixation point. Figure 13(i) shows the location of the cyclopean eye (with coordinate frame attached), a small square showing the fixated surface and, in (**A**), Eye1 and Eye2 moving up slightly (like the head bobbing up) while in (**B**) the change from Eye1 to Eye2 is a slight sideways movement. The consequence of these translations of the eye is that the epipole is just below the cyclopean fovea in (**A**) and slightly to the left of the fovea in (**B**), as shown in column (ii).

Columns (iii), (iv) and (v) show what happens when the surface is slanted. In each case, the patch is fixated, as shown in (i) and (ii) but the view is now zoomed in on the patch. In (iii), the patch faces the cyclopean point. As we have discussed above, this results in an overall expansion in the eye that is closer to the patch and a contraction in the other eye (pure 'towards' flow). In column (iv), the surface is slanted about a vertical axis. This results in a change in aspect ratio of the image projected into the cyclopean eye (black) and plus an overall expansion in Eye1 and a compression in Eye2, just as in column (iii). But, in addition to this 'towards' flow, there is a shear in the image in (**A**), in a direction parallel to epipolar lines (shown in pink) while in (**B**) there is an expansion of the image in Eye1, also in a direction parallel to epipolar lines, and a compression of the image in Eye2 in the same direction. When the axis of slant is rotated through 90 degrees in column (v) the cyclopean compression is rotated by 90 degrees and the type of flow is reversed between (**A**) and (**B**): now there is compression/expansion in (**A**) and shear in (**B**). Of course, all that has happened is that the direction of the epipolar lines has changed by 90 degrees between (**A**) and (**B**). The 'sideways' flow that gives rise to the shear or to the additional (slant-related) compression/expansion is always parallel to epipolar lines.

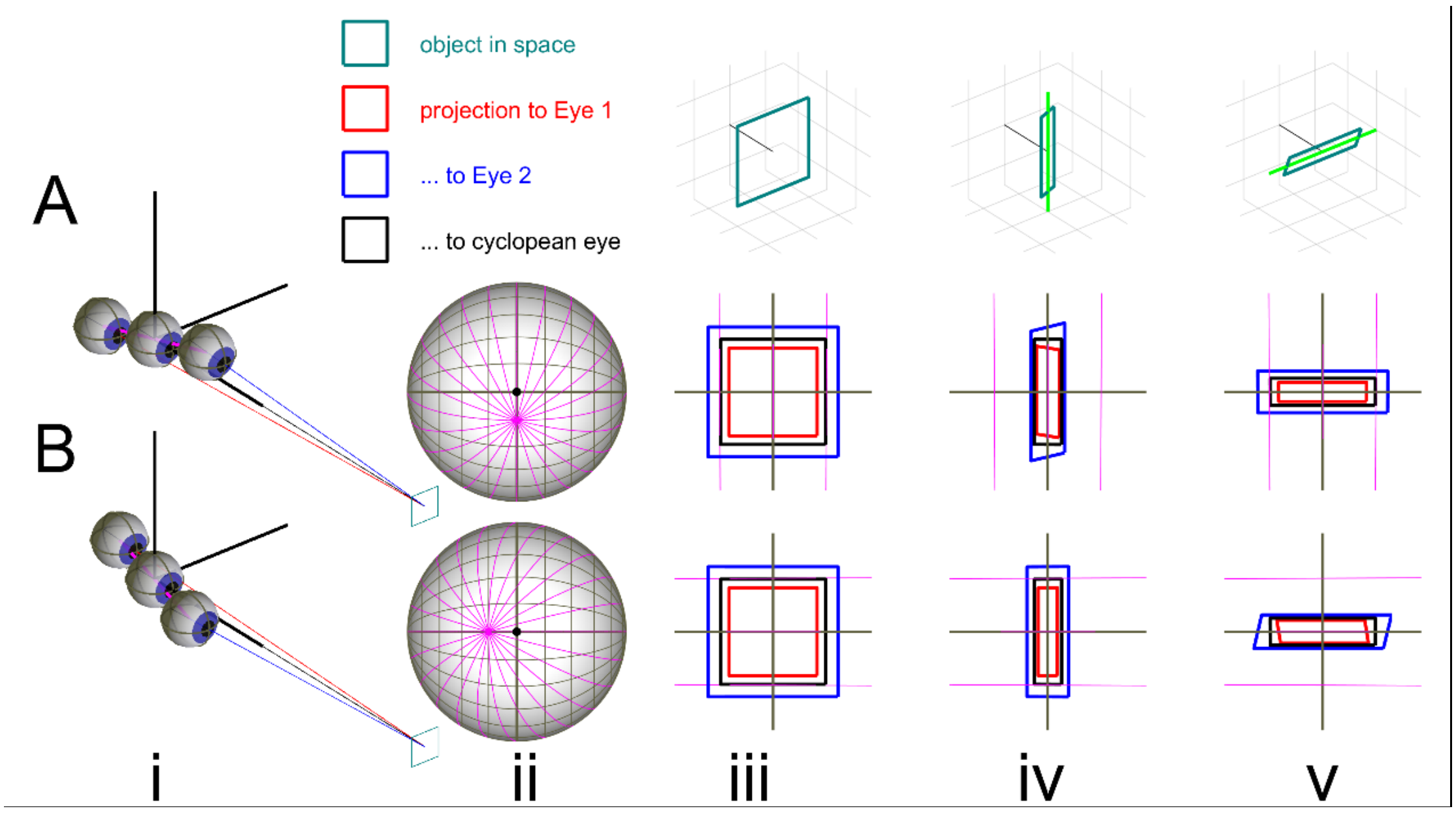

***Figure 13: The disparity or flow generated by a small planar patch near the epipole.*** *(i) Two sets of three eyes are shown on the left, arranged in a line that defines the epipolar direction – this is slightly different in the top and bottom row. All three eyes fixate the surface patch, shown as a black square that has been exaggerated in size by a factor of 30 to make it visible. In the top row, (**A**), the epipole is just below the fovea whereas in the bottom row, (**B**), the epipole is just to the left of the fovea. The columns on the left show zoomed-in pictures of the retinal projection of this patch in Eye 1, the cyclopean eye and Eye 2 (red, black, blue) when the patch is facing the cyclopean eye (iii), slanted about a vertical axis (iv), or slanted about a horizontal axis (v). The magnitude of the slant is such that in **B**(iv) and **A**(v) the patch is parallel to the line $O_1O_2$.*

Incidentally, the magnitude of the slant applied to the patch in column (iv) is such that it lies parallel to the epipolar vector in (B), ditto for column (v) in (A), i.e. in binocular terms, these are 'frontoparallel' patches (although, as we have said, it is clearly impossible for a binocular observer to see these patches through the side of the head). Numerically, the OSR for these patches is 1.477 (which is the same as the ESR and OSR in column (iii)) and the ESR is 2.182, confirming that $ESR = OSR^2$ in these cases. For the patches shown in A(iv) and B(v), OSR = ESR as expected. This is because the slant axis is in the plane of regard and so 'sideways' flow only produces a shear which does not affect ESR. If we used a retinal frame to describe the patches (which stays constant with respect to the page, as shown by the grey cross) then we would have $HSR = VSR^2$ for Figure 12 B(iv) but $VSR = HSR^2$ for Figure 13A(v).

It is important to note how extreme these conditions are compared to the situation that is usually considered for binocular vision. Read *et al* (2009) examined the relationship between horizontal size ratios and vertical size ratios using a retinal coordinate frame (with two alternative definitions of 'vertical'). They included caveats about the range of eccentric gaze (up to 15 degrees) and the range of retinal eccentricity (up to 15 degrees 'parafoveal' region) over which the relationships that they described would apply and the example shown in Figure 13 is very far outside these ranges. When the eyes of a binocular observer are in primary position, i.e. fixating 90 degrees away from the epipole, a longitude-longitude retinal coordinate frame (Figure 14A and B) is coincident with epipolar planes and in this case, by definition, HSR is the same as ESR and VSR is the same as OSR across the entire retina. The eyes can move a small amount away from primary position, either in a vertical or a horizontal direction, and the directions of the longitude-longitude retinal frame will not depart too much from epipolar and ortho-epipolar directions, at least if we restrict consideration to regions close to the fovea. Nevertheless, the situation in Figure 13 is relevant because we are considering optic flow as well as binocular disparity and, for optic flow, the pattern shown in Figure 13 is entirely typical. Also, it is important to establish the underlying geometry before considering approximations or implementations in a retinal frame.

In relation to the range of retinal eccentricities or angles of gaze over which HSR = $VSR^2$ might be a good approximation, Howard and Rogers (1995) made the following claim:

"VSRs are unaffected by either the vergence state of the eyes or their state of eccentric gaze within the plane of regard when elevation is measured according to the gun turret coordinate system" (p282)

They chose an elevation-latitude retinal coordinate frame to describe vertical position and size, so this statement is true even for large changes in vergence or gaze eccentricity. The axis of rotation for changing gaze or vergence is perpendicular to any plane of equal latitude, hence rotating about this axis does not change vertical position on the retina. However, they then go on to say:

"VSRs depend only on the relative distances of the points from the two eyes." (p282)

This is not the case (although it *is* true of OSR). The direction Howard and Rogers define as 'vertical' can differ by as much as 90 degrees from the ortho-epipolar direction, as we have seen from Figure 13, which means that 'sideways' flow can contribute to 'vertical' size. For Eqn 6 to hold, the trick is to extract a pure indicator of relative distance from the two eyes, such as 'towards' flow, and, in relation to coordinate frames, only OSR does that.

# 3. Discussion

Having set out a framework for dividing up flow into various components, we now consider how this relates to other divisions of flow (mainly binocular disparity) that have been proposed in the past. We also discuss how dividing up flow into separate components might be useful in practice.

## 3.1. Relationship to other divisions of disparity and optic flow

Table 1 lists a number of the coordinate systems that have been proposed to describe binocular disparity and how these relate to 'sideways', 'towards' and 'vergence' flow. 'Coordinate 1' and 'coordinate 2' correspond to common definitions of azimuth and elevation respectively, except for row 3 which describes a polar coordinate frame centred on the fovea.

| | Coordinate 1 | Coordinate 2 | Notes |
|---|---|---|---|
| Mayhew et al 1982, Read et al 2009; (elevation-longitude) | Mayhew et al: X<br>Read et al: α | Mayhew et al: Y<br>Read et al: η | 'Sideways-flow' ↔ constant value of Coordinate 2<br>(This coordinate frame is a longitude-longitude system) |
| Read et al 2009 (elevation-latitude), Howard and Rogers (1995) | Read et al: α<br>Howard and Rogers: α | Read et al: κ<br>Howard and Rogers: β | 'Vergence-flow' ↔ constant value of Coordinate 2 |
| Weinshall (1990), Gårding et al (1995), Glennerster et al (2001) | All: eccentricity<br>Read et al: ξ<br>Weinshall: R<br>Gårding et al, and Glennerster et al: ρ | All: Polar angle,<br>Weinshall: ν<br>Gårding et al, and Glennerster et al: θ | 'Towards-flow' ↔ constant value of Coordinate 2 |
| Helmholtz frame or 'pencil of epipolar planes' | Read et al: ζ Change in this coordinate is in 'E' direction in this paper | Read et al: λ Change in this coordinate is in 'O' direction in this paper | These coordinates relate to epipolar geometry and are independent of the orientation of the eyes. For a surface patch at the fovea, 'towards flow' is the sole contributor to change in 'O' direction |

***Table 1. Relationship between 'sideways', 'towards' and 'vergence' flow and some divisions of disparity in the literature.*** *In these cases, cyclovergence is assumed to be zero. The symbol ↔ denotes 'in the same direction as'.*

Row 1 and 2 of Table 1 both describe retinal coordinate frames with the same longitudinal system of defining eccentricity but different definitions of the vertical component (see Figure 14B and C). In Row 1, the vertical component is also defined by a longitudinal system (η). This corresponds to an (x,y) coordinate frame on a planar image, as shown in Figure 14A. It is also the direction of 'sideways' flow. On the other hand, the direction of 'vergence' flow (row 2 in Table 1) corresponds to the lines of latitude in Figure 14C (constant κ).

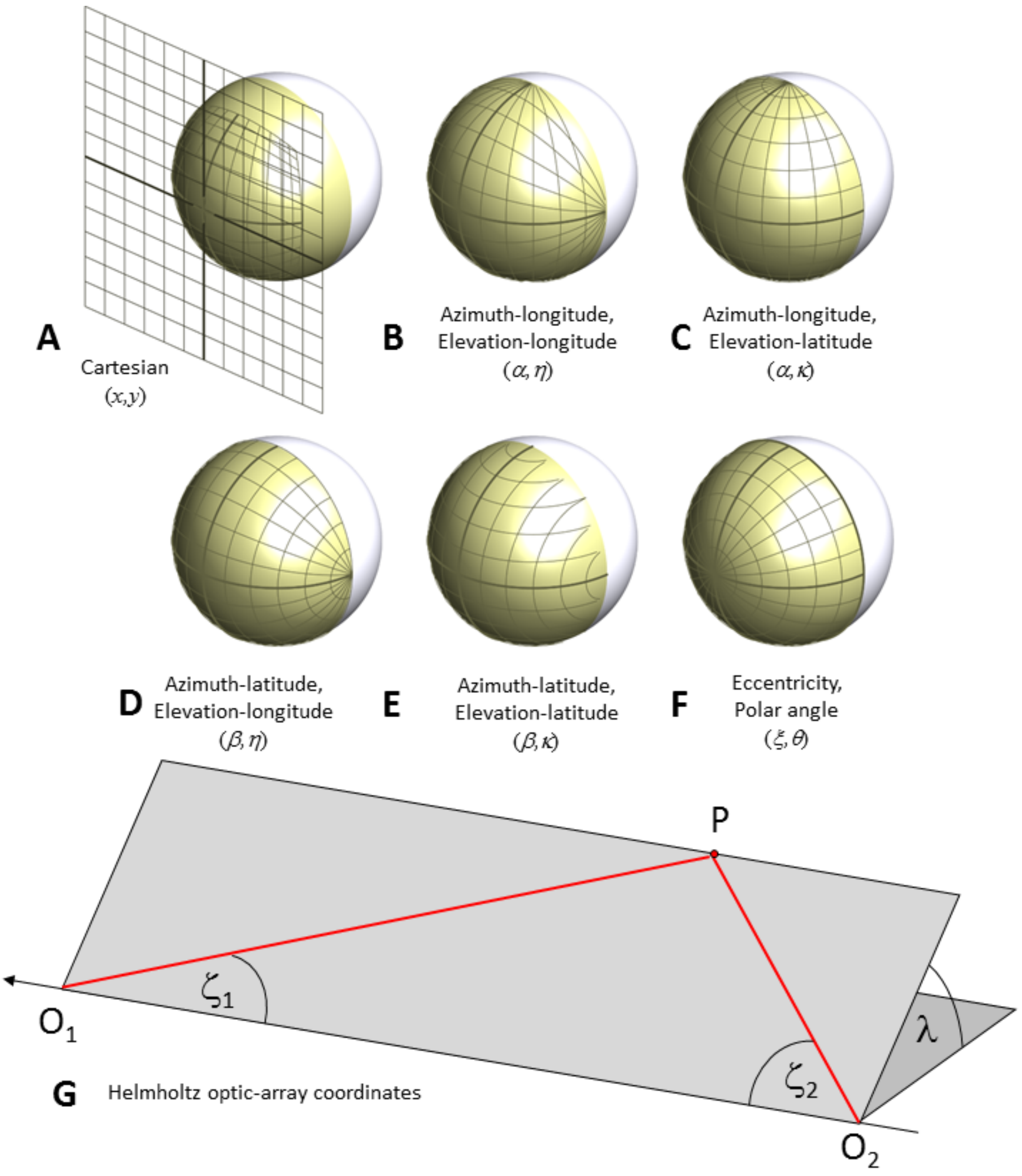

***Figure 14. Retinal coordinate frames and a Helmholtz frame.*** *This figure illustrates different possible retinal coordinate frames with the labels applied to horizontal and vertical coordinates. Horizontal is given by* $\alpha$ *in **(B)** and **(C)** or* $\beta$ *in **(D)** and **(E)**. Vertical is given by* $\eta$ *in **(B)** and**(D)** or* $\kappa$ *in **(C)** and **(E)**. Panel **(F)** shows a polar coordinate retinal frame (eccentricity,* $\xi$*, and meridional angle,* $\theta$*). Panel **(G)** shows a coordinate frame (*$\lambda$*,* $\zeta$*) that is dependent only on the location of the optic centres of the eyes, and so is not a retinal*

*coordinate frame. The planes shown in (**G**) are the same as those indicated by the flow lines in Figure 1. Symbols for coordinates correspond to those in Read et al, (2009).*

Row 3 shows a different retinal coordinate frame and, in many ways, a much more familiar one with retinal eccentricity ($\xi$) and meridional angle ($\theta$) defining a polar coordinate frame centred on the fovea. Of course, 'towards' flow corresponds to lines of constant meridional angle (e.g. expanding outwards from the fovea). The only flow that contributes to changes in meridional angle are 'sideways' flow and 'vergence' flow. We discuss some possible advantages of this coordinate frame in the next section.

The coordinate frame in row 4 comes into a different category because it is not a retinal frame. It is based simply on the epipolar geometry, i.e. the location of the two optic centres define a line at which a pencil of planes coincide. The elevation of the plane is the vertical coordinate ($\zeta$) while azimuth in the plane defines the horizontal coordinate ($\lambda$), see Figure 14G. When the two eyes are in primary position (pointing 90 degrees from the epipole), this Helmholtz coordinate system is the same as the ($\alpha$,$\eta$) or (x,y) retinal coordinate frames (Figure 14A and B) but that is no longer the case as soon as the eyes move away from primary position. The $\lambda$ and $\zeta$ coordinates measure change along and orthogonal to epipolar lines respectively. We will explore the importance of measuring flow in these directions in relation to judgements of slant in Section Flow for a small surface patch.

The literature on optic flow does not generally concern itself with the pros and cons of different retinal coordinate frames, unlike the debate in that exists for binocular disparity. Instead, optic flow in a local region has been divided into orthogonal affine components, or differential invariants, 'divergence', 'deformation' and 'curl'. These relate to the 1st order structure of the surface, i.e. local slant ('div', 'def', curl', see Koenderink (1986)). For a fixated surface, 'div' is in the same direction as 'towards' flow. 'Sideways' flow has contributions from both 'div' and 'def'. 'Curl' relates to cyclovergence flow. Koenderink points out that flow generated by rotation of the eye ('cyclovergence' and 'vergence' flow) are not influenced by surface structure. These descriptions of the components of flow do not assume any particular retinal coordinate frame.

Koenderink's analysis of optic flow applies only to a small region of the visual field. Our proposed division of translational flow into 'towards' and 'sideways' flow applies to more than just a small patch (see Figure 6 and Figure 10). In the next section, we consider the control of heading using the detection of 'towards' flow and any deviation from this (i.e. the presence of 'sideways+vergence' flow). In this example, the proposed control strategy would be most effective if it used peripheral flow.

### 3.2. Some potential uses of polar flow

The system of dividing up optic flow that we have described is not a general one; it is peculiarly biological. The definitions of 'sideways', 'towards', 'vergence' and 'cyclovergence' flow all rely on the assumption that the observer is fixating. But this assumption is an entirely valid one for humans and, indeed, most other animals, both in the case of static binocular viewing and for a moving observer. Hence, the division of optic flow and disparity that we suggest is relevant. Three examples are included below to illustrate why it might be useful: (i) controlling heading with respect to a fixated point; (ii) recovering the surface structure of a fixated surface and (iii) building up a representation of the visual direction of objects plus some information about their distance even when the head and eyes move.

For the control of heading, consider an observer moving towards a fixated object while deviating as little as possible to the left or right. It is fairly clear that for this task, a division of flow into 'towards' flow and 'sideways + vergence' flow will be helpful because 'towards' flow corresponds to achieving the task (approaching the fixated object) while any other flow ('sideways + vergence' flow) indicates that the observer has deviated off the desired path. Glennerster et al (2001) discuss this example in more detail and consider a hierarchical method for, at the simplest level, correcting heading errors and, at the most complex level, recovering the actual direction of heading. It is hierarchical because gradually adding information can gradually increase the precision with which the direction of heading is defined. This is very similar to the notion of a hierarchical recovery of surface structure that others have proposed (Tittle et al, 1995; Glennerster et al, 1996).

In relation to the recovery of surface structure, Weinshall (1990) and Gårding et al (1995) suggested that depth information about the surface could be recovered in a hierarchical manner, i.e. simple information could be recovered using a simple algorithm and then Euclidean or metric structure recovered by the addition of extra information. To do this, both papers pointed out that it would be useful to measure 'polar angle disparities' which were one component of flow (a change in the meridional angle, see row 3 of Table 1). 'Towards' flow does not change polar angle disparities (because this component is along radial lines, keeping the polar angle constant) so, by focusing on the information available from polar angle disparities they were essentially recovering information that was only available from 'sideways' flow. In particular, they showed that polar angle disparities provided useful information about the depth structure of the fixated surface (or 'relative nearness' as Gårding et al 1995 described it) – essentially, the bas relief structure or ratio of depths of points on a surface. This is useful information but falls short of full metric depth. Because polar angle disparities are invariant to the changes in 'towards' flow, they are determined only by 'sideways + vergence' flow (although they are not exactly the same: for example, polar angle disparities are zero in the plane of regard which is not true of 'sideways + vergence' flow).

Finally, in relation to building a representation that could survive head and eye movments, Glennerster et al (2001) raise a quite different reason for considering 'towards' and 'sideways + vergence' flow. The question they considered is how a moving observer might build up a representation of visual direction and approximate distance of objects surrounding them. Of course, a representation of visual direction does not involve flow, but if it is to contain information about the relative distance of objects then information from optic flow (or disparity) is required. First, Glennerster et al (2001) focus on the saccades that would take the eye between points across the entire optic array (i.e. pure rotations of the eye). Together, these saccades triangulate the whole sphere and so define the relative visual direction of objects surrounding the observer. Then, the authors consider the *changes* to this set of relative visual directions (hence optic flow) either when the observer moves or for the difference between the two eyes' views in binocular vision. Changes in the two coordinates of polar flow ($\rho$ and $\theta$ in row 3 of Table 1) provide information about the change in length of the saccades and changes in the angle between different saccades respectively. If saccade lengths and the angles between saccades are useful primitives in a primal sketch of visual direction then *changes* in $\rho$ and $\theta$ (i.e. optic flow decomposed into these two directions) are useful in building up information about the distance of objects. From this persepective, $\rho$ and θ are a particularly relevant pair of basis vectors for optic flow. For a fuller discussion of this proposal, see Glennerster et al 2001, Glennerster et al 2009 and Glennerster, 2016).

## 4. Conclusion

We have set out a general framework to describe flow that includes both binocular disparity and optic flow generated by observer (or camera) movement. We have included a description that applies to any translation and rotation of the eye (or camera), given that the translation of the eye is small compared to the distance to the scene points but we have then added constraints that are generally true for human vision such that the rotation of the eye as it translates is constrained (the eye fixates a single point and its torsion around the line of sight is such that the horizon tends to project to the same meridian on the retina). The consequence is that the types of image change (or flow) that can occur during small translations of the optic centre (including the translation from left to right eye in binocular vision) are very restricted compared to what they might be if eye rotation was not yoked to eye translation as it is in human vision. Given this restriction, we have discussed some of the advantages of separating radial flow with respect to the fovea from flow in an orthogonal direction. Overall, this treatment is more general than, and quite different from, the traditional separation of disparity into ‘horizontal’ and ‘vertical’ components.

## 5. Methods

In this section, we set out the notation for describing the relationship between world and eye coordinates and for flow as the eye translates and rotates. There is some overlap with the notation developed by Read et al (2009) for binocular vision. However, in making this treatment more general, there are also some important differences particularly in relation to the orientation of the cyclopean eye (or, for optic flow, the eye at a mid-point between time 1 and time 2).

### 5.1. Eye and world frames

Figure 7 showed the relationship between the eye-centered coordinate system ($X^e$, $Y^e$, $Z^e$) and the world-centered coordinate system ($X^w$, $Y^w$, $Z^w$). We will always include superscripts when we are mixing frames. When we omit superscripts, this indicates that any frame can be chosen so long as it applies to all vectors. For example, the dot product between two position vectors is independent of which frame is chosen to express the vectors: $\mathbf{a}.\mathbf{b} = \mathbf{a}^w.\mathbf{b}^w = \mathbf{a}^e.\mathbf{b}^e$.

When we refer to an eye's "location", we mean the location of its center of projection. Eye 1 and Eye 2 are located at $\mathbf{N}_1^w$ and $\mathbf{N}_2^w$ respectively in world-centered coordinates. We define the *epipolar vector* which points from Eye 1 to Eye 2:

$$\boldsymbol{\varepsilon} = (\mathbf{N}_2 - \mathbf{N}_1)/2.$$

The cyclopean eye is defined to be in between the two eyes:

$$\mathbf{N}_c = (\mathbf{N}_2 + \mathbf{N}_1)/2$$

In general, we therefore have $\mathbf{N}_1 = \mathbf{N}_c - \boldsymbol{\varepsilon}$ ; $\mathbf{N}_2 = \mathbf{N}_c + \boldsymbol{\varepsilon}$. In cyclopean coordinates, the cyclopean eye is at the origin: $\mathbf{N}_c^c=0$. In cyclopean coordinates, therefore, the locations of the two eyes are simply $\pm\boldsymbol{\varepsilon}^c$ .

Each eye has its own rotation matrix $R_e^w$ specifying its orientation with respect to world-centered coordinates. Given a direction in eye-centered coordinates, $\mathbf{v}^e$, to express this direction in world-centered coordinates we rotate by $R_e^w$:

$$\mathbf{v}^w = R_e^w \, \mathbf{v}^e.$$

Conversely, to express a world-centered direction in eye-centered coordinates, we rotate by $R_w^e$ , the inverse of $R_e^w$ (i.e. its transpose, since this is a rotation matrix):

$$\mathbf{v}^e = R_w^e \, \mathbf{v}^w.$$

In particular, the rotation matrix determines the direction of the eye's visual axis in world-centered coordinates, $\boldsymbol{Z}_e^w$. We defined the eye-centered coordinate system such that the visual axis is the Z-axis, so in world-centered coordinates

$$\boldsymbol{Z}_e^w = R_e^w \boldsymbol{Z}.$$

The visual axis specifies the elevation and azimuth of the eye's gaze; the torsion about the visual axis is then needed to specify the rotation matrix.

## 5.2. The pseudofixation point

Figure A1 shows the three eyes along with their visual axes, which are parallel to $\mathbf{Z}_{1,2,c}$ respectively. In general, as in this example, the visual axes do not intersect. $F$ is the pseudofixation point, defined to be the midpoint of the line joining the points where the visual axes approach most closely. This line is perpendicular to both visual axes, i.e. is parallel to $\boldsymbol{Z}_1 \times \boldsymbol{Z}_2$. We define the *vergence vector* $\boldsymbol{\delta}$ to be the vector $\boldsymbol{Z}_1 \times \boldsymbol{Z}_2$. Note that this is not a unit vector; its magnitude is equal to the sine of the angle between the visual axes, i.e. the vergence angle $H_\Delta$. Let $2d\boldsymbol{\delta}$ be the vector through F linking the visual axes; if the eyes are fixating, $d$=0.

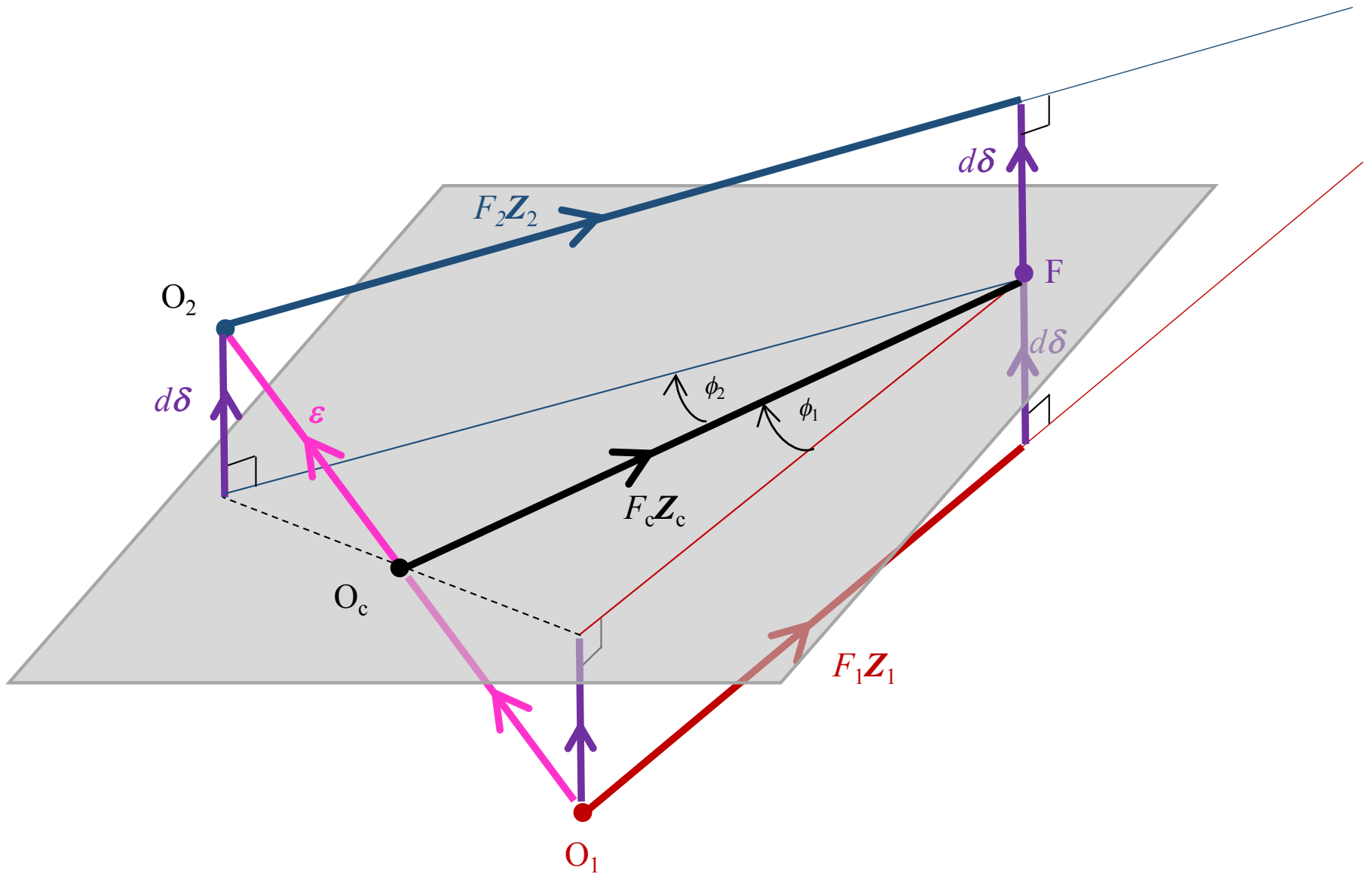

***Figure A1. Definition of a 'pseudofixation point'****. The locations of the three eyes, and the pseudofixation point F. The distance 2d|*$\boldsymbol{\delta}$*| is the minimum separation of the visual axes from Eye 1 and Eye 2; if the eyes are fixating, d=0. The shaded plane is the generalised plane of regard, formed by vectors parallel to the visual axes of the three eyes,* $\mathbf{Z}_1$*,*$\mathbf{Z}_2$ *and* $\mathbf{Z}_c$*. In this example, the visual axes starting from Eye 1 and Eye 2 (heavy red, blue lines labelled* $\mathbf{Z}_1$*,*$\mathbf{Z}_2$*) do not intersect, but when offset by the vector* $d\mathbf{Z}_1 \times \mathbf{Z}_2$ *(light lines of the same colour) they lie in the generalised plane of regard along with the vector* $\mathbf{Z}_c$ *starting from the cyclopean eye. For a general eye position, the epipolar vector* $\boldsymbol{\varepsilon}$ *(pink) does not lie in the generalised plane of regard. If the eyes are fixating, then d=0 and* $\boldsymbol{\varepsilon}$ *does also lie in this plane.* $F_{1,2,c}$ *are the distances from each eye to the point on its visual axis which is closest to the pseudofixation point. We have not indicated the frame, since the same relationships hold whether the vectors are expressed in world-centered coordinates,* $\mathbf{Z}_1^w$*,*$\mathbf{Z}_2^w$*,* $\mathbf{Z}_c^w$ *and* $\boldsymbol{\varepsilon}^w$*, or cyclopean-eye centered coordinates* $\mathbf{Z}_1^c$*,*$\mathbf{Z}_2^c$*,* $\mathbf{Z}$ *and* $\boldsymbol{\varepsilon}^c$*.*

We define $F_{1,2,c}$ to be the distance from each eye along its visual axis to the point of closest approach to the pseudofixation point F. Examining Figure 14, we see that

$$F_c \boldsymbol{Z}_c = \boldsymbol{\varepsilon} + F_2 \boldsymbol{Z}_2 - d\,\boldsymbol{\delta} = -\,\boldsymbol{\varepsilon} + F_1 \boldsymbol{Z}_1 + d\,\boldsymbol{\delta}$$

Eq A1

We can solve Eq A1 to obtain $\boldsymbol{Z}_c$. Taking the inner product of both sides with $\boldsymbol{Z}_1 \times \boldsymbol{Z}_2$, and using the fact that $\boldsymbol{Z}_1.(\boldsymbol{Z}_1 \times \boldsymbol{Z}_2) = \boldsymbol{Z}_2.(\boldsymbol{Z}_1 \times \boldsymbol{Z}_2) = 0$, we obtain

$$F_c \boldsymbol{Z}_c.\boldsymbol{\delta} = \boldsymbol{\varepsilon}.\boldsymbol{\delta} - d = -\,\boldsymbol{\varepsilon}.\boldsymbol{\delta} + d\,.$$

Eq A2

This equation states that $F_c \boldsymbol{Z}_c.\, \boldsymbol{\delta}$ is equal to its own negative, so must be zero. Since the fixated object is not inside the eye, $F_c > 0$ and so

and thus

$$\boldsymbol{Z}_c.\, \boldsymbol{\delta} = \boldsymbol{Z}_c.\, (\boldsymbol{Z}_1 \times \boldsymbol{Z}_2) = 0.$$

That is, with our definition of the cyclopean eye, the visual axes of all three eyes lie in a plane, the *generalised plane of regard*, shown shaded in Figure 14. This definition is fully general and does not require the eyes to be fixating on a single point in space.

If the eyes *are* fixating on a single point, then the distance $d$ in Figure 14 is zero. It follows from Eq A2 that

$\boldsymbol{\varepsilon}.\boldsymbol{\delta} = 0$ for a fixating eye posture

i.e. for fixation, the axis of vergence must lie at 90 degrees from the epipole. In the body of the paper, we have referred to the plane of regard as plane $O_1O_2F$. This remains true and is one special case of the generalised plane of regard defined here, i.e. with $d = 0$.

## 5.3. Deriving the cyclopean eye's rotation matrix

We have chosen to make the cyclopean eye look directly at the pseudofixation point. This fixes the azimuth and elevation of the cyclopean eye. As we stated in the main text, we define its torsion to be exactly in between that of Eye 1 and Eye 2. We now go through how this works.

### *5.3.1. Deriving the cyclopean eye's rotation matrix given the locations and orientations of Eye 1 and Eye 2 in world-centered coordinates*

To begin with, we assume we are given $\mathbf{N}_1^w$ and $N_2^w$ and the rotation matrices $R_1^w$ and $R_2^w$, which specify how the two physical eyes are positioned with respect to world-centered coordinate axes. The location of the cyclopean eye in world-centered coordinates is easy: $\mathbf{N}_c^w = (\mathbf{N}_2^w+\mathbf{N}_1^w)/2$. We now explain how we define its rotation matrix, $R_c^w$.

The rotation matrix $R_1^w$ gives us the coordinate axes of Eye 1 expressed in world-centered coordinates. For example, the visual axis of Eye 1, expressed in world-centered coordinates, is:

$$\boldsymbol{Z}_1^w = R_1^w \boldsymbol{Z}$$

Conversely, the inverse rotation, $R_w^1$ , rotates the visual axis of Eye 1, expressed in world-centered coordinates, back onto the world-centered Z axis:

$$R_w^1 \boldsymbol{Z}_1^w = \boldsymbol{Z}$$

For Eye 2, similarly

$$\boldsymbol{Z}_2^w = R_2^w \boldsymbol{Z}$$

and so $\boldsymbol{Z}_2^w = R_2^w R_w^1 \boldsymbol{Z}_1^w$.

The rotation matrix $R_2^w R_w^1$ rotates the coordinate frame of Eye 1 onto the coordinate frame of Eye 2, when both frames are expressed in world-centered coordinates (Fig A2A). If there were no cyclovergence between the eyes, $R_2^w R_w^1$ would be a pure vergence rotation about $\boldsymbol{\delta}^w$ :

$$R_2^w R_w^1 = \mathrm{Rot}[\mathbf{Z}_1^w \times \mathbf{Z}_2^w] = \mathrm{Rot}[\boldsymbol{\delta}^w] \quad \text{(for zero cyclovergence).}$$

where we have introduced the notation that Rot[**v**] is a rotation matrix defining a rotation through angle arcsin(|**v**|) about vector **v**. $\mathbf{Z}_1 \times \mathbf{Z}_2$ is parallel to the vergence axis $\boldsymbol{\delta}$, and its magnitude is equal to the sine of the vergence angle between the visual axes of Eye 1 and Eye 2, $H_\Delta$ (cyan rotation in Fig A2B).

However, in general, the eyes will also be cycloverged relative to one another and therefore we also need to add a rotation around the visual axis through the cyclovergence angle $T_\Delta$ (gold rotation in Fig A2B). It doesn't matter whether we first cycloverge through $T_\Delta$ about visual axis 1 and then do the vergence from Eye 1 to Eye 2, as shown in Fig A2B:

$$R_2^w R_w^1 = \mathrm{Rot}[\boldsymbol{\delta}^w]\mathrm{Rot}[\mathbf{Z}_1^w \sin T_\Delta]$$

or whether we first verge from Eye 1 to Eye 2 and then cycloverge about visual axis 2:

$$R_2^w R_w^1 = \mathrm{Rot}[\mathbf{Z}_2^w \sin T_\Delta]\mathrm{Rot}[\boldsymbol{\delta}^w]$$

Eq A3

or whether we verge from Eye 1 to the cyclopean eye, cycloverge about the cyclopean visual axis, and then continue the vergence onto Eye 2:

$$R_2^w R_w^1 = \mathrm{Rot}[\mathbf{Z}_c^w \times \mathbf{Z}_2^w]\ \mathrm{Rot}[\mathbf{Z}_c^w \sin(T_\Delta)]\ \mathrm{Rot}[\mathbf{Z}_1^w \times \mathbf{Z}_c^w]$$

A rotation about axis 1 followed by a rotation about axis 2 is identical to a rotation about axis 1, followed by a rotation about the new axis 2:

$$\mathrm{Rot}[\mathbf{v}_2]\mathrm{Rot}[\mathbf{v}_1] = \mathrm{Rot}[\ \mathrm{Rot}[\mathbf{v}_2]\ \mathbf{v}_1\ ]\ \mathrm{Rot}[\mathbf{v}_2]$$

Using this identity, it is straightforward to confirm that the three alternative expressions for $R_2^w R_w^1$ are the same. We now need to solve for the cyclovergence angle. Taking Eq A3, for example, we have

$$\mathrm{Rot}[\mathbf{Z}_2^w \sin T_\Delta] = R_2^w R_w^1 \, \mathrm{Rot}[-\boldsymbol{\delta}^w]$$

All the quantities in the right-hand side are known, so it is easy to solve for the cyclovergence $T_\Delta$. Using the fact that the trace of a rotation matrix is 1 plus twice the cosine of the rotation angle, we have

$$2\cos T_\Delta = \mathrm{Tr}(\, R_2^w R_w^1 \, \mathrm{Rot}[-\boldsymbol{\delta}^w] \,) - 1$$

Eq A4

We are now finally in a position to define the rotation matrix of the cyclopean eye. the rotation matrix $R_c^w R_w^1$ rotates the coordinate frame of Eye 1 onto the coordinate of the cyclopean eye, while the rotation matrix $R_2^w R_w^c$ rotates the coordinate frame of the cyclopean eye onto the coordinate frame of Eye 2 (all expressed in world-centered coordinates).

As noted, we wish to define the cyclopean coordinates such that they are exactly midway between the coordinate frame of the two eyes. Thus, to get to cyclopean coordinates from Eye 1, we first do a vergence rotation through $\phi_1$, and then rotate through half the cyclovergence angle, which we call $T_\delta$ in accordance with the notation of Read et al. (2009).

$$R_c^w R_w^1 = \mathrm{Rot}[\mathbf{Z}_c^w \sin(T_\delta)] \, \mathrm{Rot}[\mathbf{Z}_1^w \times \mathbf{Z}_c^w]$$

Then, to get to Eye 2 from cyclopean coordinates, we rotate through the other half of the cyclovergence angle and then do a vergence rotation through $\phi_2$:

$$R_2^w R_w^c = \mathrm{Rot}[\mathbf{Z}_c^w \times \mathbf{Z}_2^w] \, \mathrm{Rot}[\mathbf{Z}_c^w \sin(T_\delta)]$$

The cyclopean rotation matrix is then easily solved for:

$$R_c^w = \mathrm{Rot}[\mathbf{Z}_c^w \sin(T_\delta)] \, \mathrm{Rot}[\mathbf{Z}_1^w \times \mathbf{Z}_c^w] \, R_1^w$$

or equivalently

$$R_c^w = \mathrm{Rot}[-\mathbf{Z}_c^w \sin(T_\delta)]\ \mathrm{Rot}[\mathbf{Z}_2^w \times \mathbf{Z}_c^w]\ R_2^w$$

Eq A5

Thus, we have now produced a complete definition of the cyclopean eye for any eye postures $R_1^w$ and $R_2^w$.

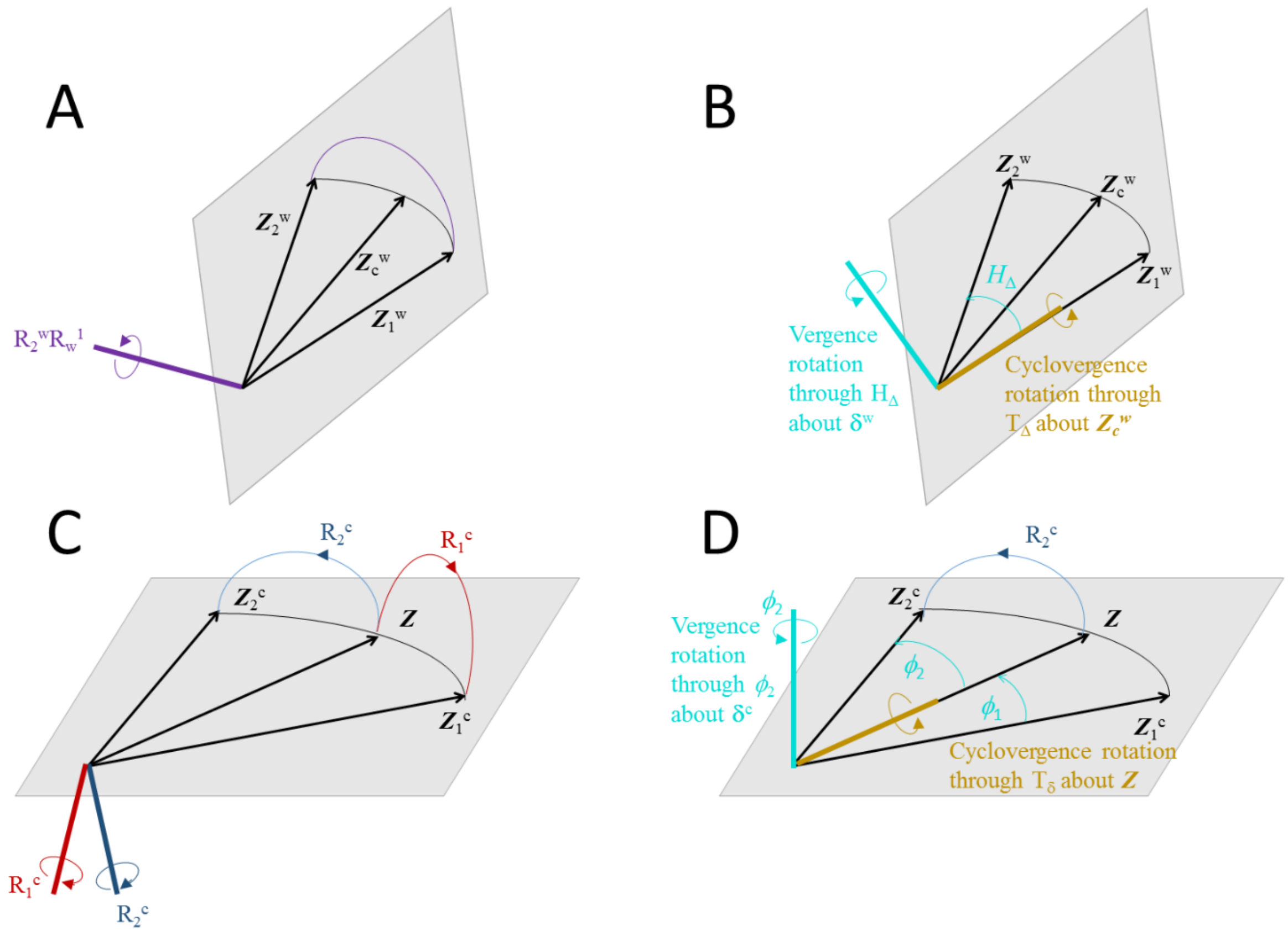

***Fig A2. How the rotation from Eye 1 to Eye 2 can be decomposed into a vergence and a torsion component.*** *In each panel, the shaded plane indicates the plane of regard. AB: In world-centered coordinates; CD: in cyclopean coordinates. A: Rotation matrix $R_2^w R_w^1$ maps the coordinate axes of Eye 1 into the coordinate axes of Eye 2. In general, this rotation is about an arbitrary axis, shown by the heavy purple line. B: This can be decomposed into a vergence rotation through angle $H_\Delta = \phi_1 + \phi_2$ about the vergence axis $\boldsymbol{\delta}^w$, plus a*

*cyclovergence rotation through angle $T_{\Delta}$ about the visual axis. C: Matrix $R_1^c$ maps the cyclopean eye to Eye 1, and $R_c^2$ maps the cyclopean eye to Eye 2. In general, these rotations are about different axes, shown with the red,blue heavy lines. D: Shows how $R_2^c$ can be decomposed into a cyclovergence component T, which is a rotation through $T_{\delta}$ about the cyclopean visual axis, followed by a vergence rotation $V_c^2$ through $\phi_2$ about the vergence axis $\delta^c$. Similarly, $R_1^c$ can be decomposed into a cyclovergence rotation $T^T$ through $-T_{\delta}$ about the cyclopean visual axis, followed by a vergence rotation $V_c^1$ through $-\phi_1$ about the vergence axis (Eq A6).*

### 5.3.2. Deriving cyclopean rotation matrices for Eye 1 and Eye 2, given cyclopean visual axes and cyclovergence

For many applications, we do not care about where the eyes are located in a world-centered coordinate system, but simply want the rotation matrices relating Eye 1 and Eye 2 to the cyclopean eye. We can get these simply from Eq A5 by defining the world-centered frame to be equal to the cyclopean frame. $R_c^w$ is then the identity matrix, $\mathbf{Z}_c^w$ is simply the Z axis. We then have

$$R_c^1 = \mathrm{Rot}[\mathbf{Z}\sin T_{\delta}]\,\mathrm{Rot}[\mathbf{Z}_1^c \times \mathbf{Z}]\,; \qquad R_1^c = \mathrm{Rot}[\mathbf{Z} \times \mathbf{Z}_1^c]\,\mathrm{Rot}[-\mathbf{Z}\sin T_{\delta}]$$

$$R_c^2 = \mathrm{Rot}[-\mathbf{Z}\sin T_{\delta}]\mathrm{Rot}[\mathbf{Z}_2^c \times \mathbf{Z}]\,; \qquad R_2^c = \mathrm{Rot}[\mathbf{Z} \times \mathbf{Z}_2^c]\,\mathrm{Rot}[\mathbf{Z}\sin T_{\delta}]$$

Eq A6

We define the cyclovergence rotation matrix

$$T = \mathrm{Rot}[\mathbf{Z}\sin T_{\delta}]$$

Eq A7

and two different vergence rotation matrices, both about the vergence axis $\boldsymbol{\delta}^c$ but through

different angles:

$$V_c^1 = \mathrm{Rot}[\mathbf{Z}_1^c \times \mathbf{Z}]; \qquad V_c^2 = \mathrm{Rot}[\mathbf{Z}_2^c \times \mathbf{Z}]$$

Eq A8

As an example, suppose we are told (i) the eyes are fixating; (ii) the distance of the fixation point $F_c$ from the cyclopean eye, (iii) the epipolar vector in cyclopean coordinates, $\boldsymbol{\varepsilon}^c$, and (iv) the cyclovergence $T_\Delta$.

From this information, we can deduce the direction of vergence vector $\boldsymbol{\delta}^c$, since it is perpendicular to both $\mathbf{Z}$ and (for a fixating eye posture) to the epipolar vector $\boldsymbol{\varepsilon}^c$. From Eq A1 with $d$=0 for fixation, we have

$$F_c \boldsymbol{Z} = \boldsymbol{\varepsilon}^c + F_2 \boldsymbol{Z}_2^c = -\boldsymbol{\varepsilon}^c + F_1 \boldsymbol{Z}_1^c$$

This gives us the direction of both visual axes:

$$F_2 \boldsymbol{Z}_2^c = F_c \boldsymbol{Z} - \boldsymbol{\varepsilon}^c \; ; \qquad F_1 \boldsymbol{Z}_1^c = F_c \boldsymbol{Z} + \boldsymbol{\varepsilon}^c$$

Since $\boldsymbol{Z}_{1,2,c}$ are unit vectors, this also gives us the distance from each eye to the fixation point:

$$F_1 = |F_c \boldsymbol{Z} + \boldsymbol{\varepsilon}^c| \; ; \qquad F_2 = |F_c \boldsymbol{Z} - \boldsymbol{\varepsilon}^c|$$

and thus the direction of each visual axis:

$$\boldsymbol{Z}_1^c = (F_c \boldsymbol{Z} + \boldsymbol{\varepsilon}^c)/F_1; \quad \boldsymbol{Z}_2^c = (F_c \boldsymbol{Z} - \boldsymbol{\varepsilon}^c)/F_2.$$

We can then solve Eq A6 to obtain the rotation matrices of the two eyes in this cyclopean frame.

For some applications, one may wish to define the horizontal meridian of the cyclopean eye to be the plane of regard. In this case, simply follow the scheme above but with $\boldsymbol{\varepsilon}^c$ parallel to $\mathbf{X}$.

## 5.4. Relationship to Helmholtz coordinates for binocular vision

In binocular vision, the "world-centered" coordinate system will usually be head-centered. Here, we consider the head-centered coordinate system of Read et al (2009). Then Eye 1 is the left eye and Eye 2 the right.

$$\boldsymbol{\epsilon}^w = -\frac{I}{2}\begin{pmatrix}1\\0\\0\end{pmatrix};\ \boldsymbol{N}_1^w = -\boldsymbol{\epsilon}^w = \frac{I}{2}\begin{pmatrix}1\\0\\0\end{pmatrix};\ \boldsymbol{N}_2^w = \boldsymbol{\epsilon}^w = -\frac{I}{2}\begin{pmatrix}1\\0\\0\end{pmatrix};\ \boldsymbol{N}_c^w = \begin{pmatrix}0\\0\\0\end{pmatrix}$$

The head-centered rotation matrices are

$$R_e^w = \begin{bmatrix}1 & 0 & 0\\ 0 & \cos V_e & -\sin V_e\\ 0 & \sin V_e & \cos V_e\end{bmatrix}\begin{bmatrix}\cos H_e & 0 & \sin H_e\\ 0 & 1 & 0\\ -\sin H_e & 0 & \cos H_e\end{bmatrix}\begin{bmatrix}\cos T_e & -\sin T_e & 0\\ \sin T_e & \cos T_e & 0\\ 0 & 0 & 1\end{bmatrix}$$

where $V_e$, $H_e$, $T_e$ are respectively the Helmholtz elevation, azimuth and torsion for Eye e (see below for the definitions of these for the cyclopean eye). The visual axis for Eye e points in the direction

$$\boldsymbol{Z}_e^w = \begin{pmatrix}\sin H_e\\ -\sin V_e\cos H_e\\ \cos V_e\cos H_e\end{pmatrix}$$

For fixating eye postures, the Helmholtz elevation must be the same for all eyes: $V_e=V$. Then the vergence axis is

$$\boldsymbol{\delta}^w = \boldsymbol{Z}_1^w \times \boldsymbol{Z}_2^w = \sin H_\Delta \begin{pmatrix}0\\ \cos V\\ \sin V\end{pmatrix}$$

That is, for fixating eye postures, the vergence rotation is a rotation through the vergence angle $H_\Delta$ about an axis tilted forward from the Y axis by the Helmholtz elevation angle V.

The cyclovergence, defined in Eq A4, is $T_\Delta=(T_2-T_1)$, i.e. the difference between the Helmholtz torsions for right and left eye. The cyclotorsion of the cyclopean eye is $T_c=(T_2+T_1)/2$, the mean of the Helmholtz torsions, as defined in Read et al (2009). The vergence angle is $H_\Delta=(H_2-H_1)$, again as in Read et al (2009). However, note that we here define the azimuth of the cyclopean eye differently. In Read et al (2009), we defined the cyclopean azimuth to be the mean of the Helmholtz azimuths for right and left eye. However, this means that the visual axis of the cyclopean eye does not point at the fixation point. In order to achieve that, we here define the cyclopean azimuth slightly differently:

$$H_c = H_1 + \phi_1 = H_2 - \phi_2$$

where the angles $\phi$ are defined in Figure A2 and Figure A3.

$$\sin\theta_1 = \frac{\cos H_1 \sin H_\Delta}{\sqrt{\sin^2(H_1 + H_2)\ +4\cos^2 H_1 \cos^2 H_2}}$$

$$\sin\theta_2 = \frac{\cos H_2 \sin H_\Delta}{\sqrt{\sin^2(H_1 + H_2)\ +4\cos^2 H_1 \cos^2 H_2}}$$

This means that

$$H_c = (H_1 + H_2 + \phi_1 - \phi_2) / 2$$

differing by $(\phi_1 - \phi_2)/2$ from the gaze angle of the cyclopean eye defined in the earlier paper.

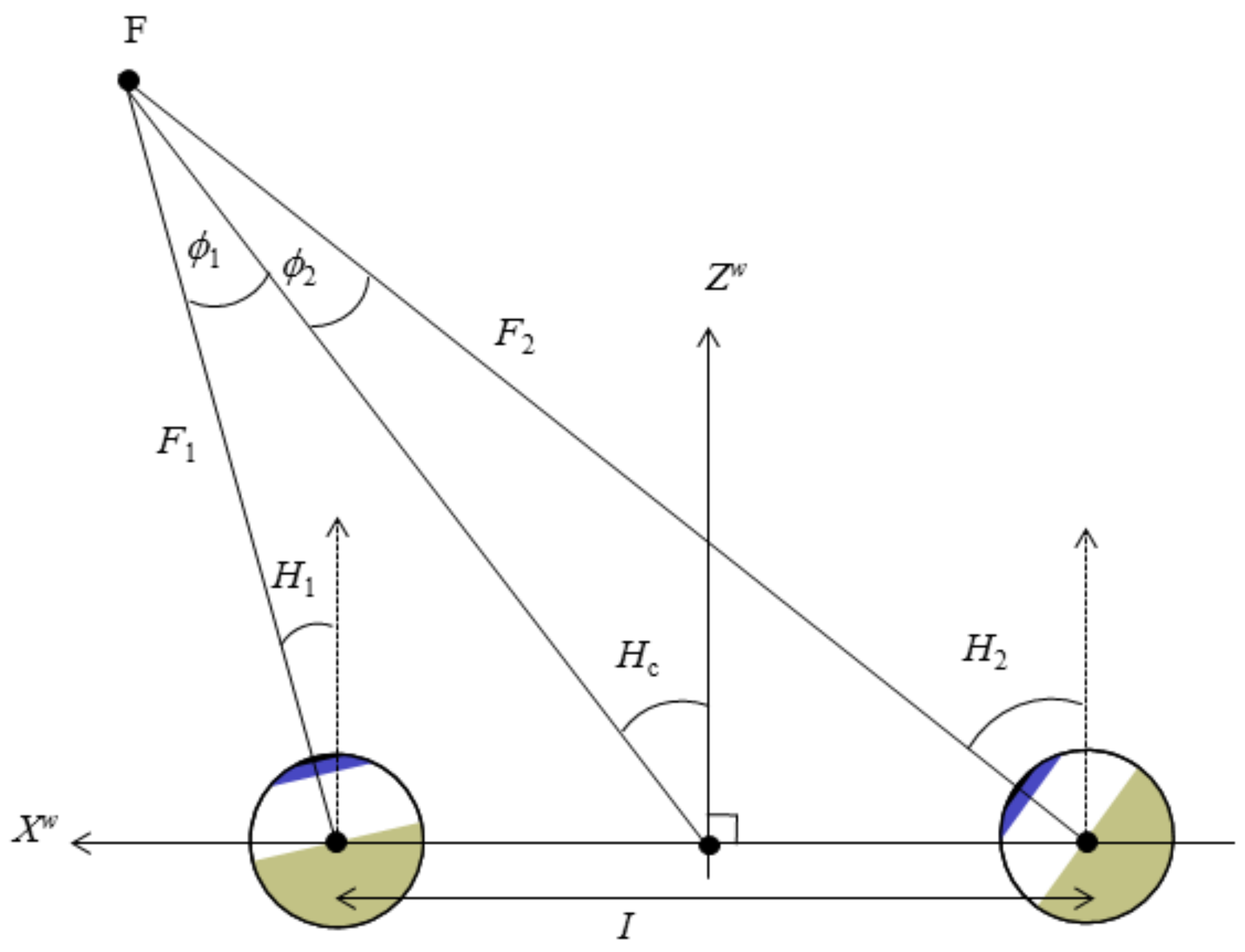

***Figure A3. Definition of vergence angle and gaze azimuth.*** *The vergence angle is* $H_\Delta = H_2 - H_1 = \phi_2 + \phi_1$. *The cyclopean gaze azimuth is* $H_c = H_1 + \phi_1 = H_2 - \phi_2$.

## 5.5. Useful relationships

Here we collect together for reference the terms defined in this paper and some useful relationships between them.

### *5.5.1. Vectors*

All vectors can be given superscripts to indicate which coordinate system they are defined in.

$\mathbf{N}_1$, $\mathbf{N}_2$ : location (i.e. center of projection) of Eye 1, Eye 2 respectively.

$\mathbf{N}_c = (\mathbf{N}_2+\mathbf{N}_1)/2$ : location of the cyclopean eye

$\boldsymbol{\varepsilon} = (\mathbf{N}_2-\mathbf{N}_1)/2$ : the epipolar vector. This points in the direction of the epipole, and we

define it as half the vector linking Eye 1 and Eye 2.

$\mathbf{Z}_1, \mathbf{Z}_2, \mathbf{Z}_c$: unit vectors along the visual axis of the respective eye.

$\boldsymbol{\delta} = \boldsymbol{Z}_1 \times \boldsymbol{Z}_2$ : the vergence vector, pointing along the axis for vergence rotations. Note that $\boldsymbol{\delta}.\boldsymbol{Z}_c=0$. For fixation, $\boldsymbol{\varepsilon}.\boldsymbol{Z}_c=0$.

**P**: an example scene point.

### 5.5.2. Rotation matrices

$R_c^w$: The rotation matrix expressing the coordinate frame of the cyclopean eye in world-centered coordinates (and similarly for $R_1^w, R_2^w$)

$\boldsymbol{Z}_c^w = R_c^w \boldsymbol{Z}$ : The visual axis of the cyclopean eye in world-centered coordinates (and similarly $\boldsymbol{Z}_1^w = R_1^w \boldsymbol{Z}$; $\boldsymbol{Z}_2^w = R_2^w \boldsymbol{Z}$).

$R_w^c = (R_c^w)^T$: The rotation matrix expressing the world-centered coordinate frame in the coordinates of the cyclopean eye.

$R_1^c$ : The rotation matrix expressing the coordinate frame of Eye 1 in cyclopean coordinates (and similarly for $R_2^c$).

$\boldsymbol{Z}_1^c = R_1^c \boldsymbol{Z}$ : The visual axis of Eye 1 in cyclopean coordinates.

$\boldsymbol{Z}_2^c = R_2^c \boldsymbol{Z}$ : The visual axis of Eye 2 in cyclopean coordinates.

$\boldsymbol{Z}_1^w = R_1^w \boldsymbol{Z}$ : The visual axis of Eye 1 in world-centered coordinates.

$\boldsymbol{Z}_2^w = R_2^w \boldsymbol{Z}$ : The visual axis of Eye 2 in world-centered coordinates.

$R_1^w = (R_w^1)^T$

$R_2^w = (R_w^2)^T$

$$R_1^c = (R_c^1)^T = R_w^c R_1^w = \mathrm{Rot}[\mathbf{Z} \times \mathbf{Z}_1^c]\, \mathrm{Rot}[-\mathbf{Z}\sin T_\delta] = \mathrm{Rot}[-\boldsymbol{\delta}\hat{}^c \sin\phi_1]\, \mathrm{Rot}[-\mathbf{Z}\sin T_\delta]$$

$$R_2^c = (R_c^2)^T = R_w^c R_2^w = \mathrm{Rot}[\mathbf{Z} \times \mathbf{Z}_2^c]\, \mathrm{Rot}[\mathbf{Z}\sin T_\delta] = \mathrm{Rot}[+\boldsymbol{\delta}\hat{}^c \sin\phi_2]\, \mathrm{Rot}[+\mathbf{Z}\sin T_\delta]$$

Eq_RotationMatrices

### 5.5.3. Mapping between visual axes

$\boldsymbol{Z}_1{}^{w} = R_1{}^{w}\boldsymbol{Z} = R_1{}^{w}R_c{}^{1}\boldsymbol{Z}_1{}^{c} = R_1{}^{w}R_w{}^{2}\,\mathbf{Z}_2{}^{w} = R_1{}^{w}R_w{}^{c}\,\mathbf{Z}_c{}^{w}$

$\boldsymbol{Z}_2{}^{w} = R_2{}^{w}\boldsymbol{Z} = R_2{}^{w}R_c{}^{2}\boldsymbol{Z}_2{}^{c} = R_2{}^{w}R_w{}^{1}\,\mathbf{Z}_1{}^{w} = R_2{}^{w}R_w{}^{c}\,\mathbf{Z}_c{}^{w}$

$\boldsymbol{Z}_c{}^{w} = R_c{}^{w}\boldsymbol{Z} = R_c{}^{w}R_w{}^{1}\boldsymbol{Z}_1{}^{w} = R_c{}^{w}R_w{}^{2}\boldsymbol{Z}_2{}^{w}$

$\boldsymbol{Z}_1{}^{c} = R_1{}^{c}\boldsymbol{Z} = R_1{}^{c}R_c{}^{2}\boldsymbol{Z}_2{}^{c}$

$\boldsymbol{Z}_2{}^{c} = R_2{}^{c}\boldsymbol{Z} = R_2{}^{c}R_c{}^{1}\boldsymbol{Z}_1{}^{c}$

### 5.5.4. Angles

$H_\Delta$ is the vergence angle, i.e. the angle between the visual axes $\boldsymbol{Z}_1$ and $\boldsymbol{Z}_2$. This is consistent with the definition in Read et al (2009).

In our definition, the magnitude of the vergence vector is the sine of the vergence angle:

$$|\boldsymbol{\delta}| = \sin H_\Delta$$

We define $\phi_{1,2}$ to be the angle between the cyclopean visual axis and the visual axis of Eye 1, 2. The sign is defined by

$$\boldsymbol{Z}_1 \times \boldsymbol{Z}_c = \hat{\boldsymbol{\delta}}\sin\phi_1$$

$$\boldsymbol{Z}_c \times \boldsymbol{Z}_2 = \hat{\boldsymbol{\delta}}\sin\phi_2$$

where $\hat{\boldsymbol{\delta}}$ is the unit vector parallel to $\boldsymbol{\delta}$.

In terms of these angles:

$$H_\Delta = \phi_1+\phi_2$$

In terms of the gaze azimuths in Helmholtz coordinates for binocular vision, as defined in Read et al. (2009)

$$H_\Delta = H_2-H_1.$$

The Helmholtz gaze azimuth of the cyclopean eye is

$$H_c = (H_1+ H_2 + \phi_1 - \phi_2) / 2$$

(slightly different from the definition $(H_1+H_2)/2$ used in Read et al (2009).

$T_\Delta$ is the cyclovergence between Eye 1 and Eye 2. In terms of the torsion in Helmholtz coordinates for binocular vision, as defined in Read et al. (2009)

$$T_\Delta=(T_2-T_1).$$

### 5.5.5. Converting between coordinate systems

We only use the epipolar and vergence vectors as directions, so to convert between world-centered and cyclopean frames, we simply rotate:

$$\boldsymbol{\varepsilon}^c = R_w{}^c\boldsymbol{\varepsilon}^w \quad ; \quad \boldsymbol{\delta}^c = R_w{}^c\boldsymbol{\delta}^w$$

$$\boldsymbol{\varepsilon}^w = R_c{}^w\boldsymbol{\varepsilon}^c \quad ; \quad \boldsymbol{\delta}^w = R_c{}^w\boldsymbol{\delta}^c$$

When projecting scene points into the eyes, we also need to take into account the translation offset between frames:

$$\mathbf{P}^1 = R_w{}^1\,(\mathbf{P}^w - \mathbf{N}_1{}^w) = R_c{}^1\,(\boldsymbol{P}^c + \boldsymbol{\varepsilon}^c)$$

$$\mathbf{P}^2 = R_w{}^2\,(\mathbf{P}^w - \mathbf{N}_2{}^w) = R_c{}^2\,(\boldsymbol{P}^c - \boldsymbol{\varepsilon}^c)$$

$$\mathbf{P}^c = R_w{}^c\,(\mathbf{P}^w - \mathbf{N}_c{}^w)$$

where $\mathbf{P}^w$ gives the world-centered coordinates of a scene point P (i.e. the world-centered vector from the world-centered origin to the scene point), and $\mathbf{P}^e$ gives the coordinates relative to Eye e (i.e. the eye-centered vector from the nodal point of that eye to the scene point).

## Acknowledgements

Supported by EPSRC EP/K011766/1 and EP/N019423/1 to AG and Leverhulme Trust Research Leadership Award RL-2012-019 to JCAR.